\documentclass[pra,twocolumn,showpacs,amsmath,superscriptaddress,preprintnumbers,amssymb,10pt]{revtex4-1}
\usepackage{graphicx}
\usepackage{dcolumn}
\usepackage{bm}
\usepackage{color}
\usepackage{subfigure}
\begin{document}

\newcommand{\SP}[1]{{\color{blue} [[#1. Saverio]]}}

\def\e{\mathrm{e}}
\def\ii{\mathrm{i}}
\def\d{\mathrm{d}}

\title{Split and overlapped binary solitons in optical lattices}

\author{Golam Ali Sekh}
\affiliation{Istituto Nazionale di Fisica Nucleare (INFN), Sezione
di Bari, I-70126 Bari, Italy}
\affiliation{Department of Physics,
University of Kashmir, Hazratbal, Srinagar-190006, J $\&$ K,
India}
\author{Francesco V. Pepe}
\affiliation{Dipartimento di Fisica and MECENAS, Universit\`a di Bari, I-70126 Bari, Italy}
\affiliation{Istituto Nazionale di Fisica Nucleare (INFN), Sezione di Bari, I-70126 Bari, Italy}
\author{Paolo Facchi}
\affiliation{Dipartimento di Fisica and MECENAS, Universit\`a di Bari, I-70126 Bari, Italy}
\affiliation{Istituto Nazionale di Fisica Nucleare (INFN), Sezione di Bari, I-70126 Bari, Italy}
\author{Saverio Pascazio}
\affiliation{Dipartimento di Fisica and MECENAS, Universit\`a di Bari, I-70126 Bari, Italy}
\affiliation{Istituto Nazionale di Fisica Nucleare (INFN), Sezione di Bari, I-70126 Bari, Italy}
\author{Mario Salerno}
\affiliation{Dipartimento di Fisica ``E. R. Caianiello",  INFN-
Gruppo collegato di Salerno, and CNISM, Universit\`a di Salerno,
I-84084 Fisciano, Italy}

\begin{abstract}
We analyze the energetic and dynamical properties of bright-bright (BB)
soliton pairs in a binary mixture of Bose-Einstein condensates subjected to the action of a combined optical lattice, acting as an external potential for the first
species, while modulating the intraspecies coupling constant of
the second. In particular, we use a variational approach and direct numerical integrations  to investigate the existence and stability  of BB solitons in which the two species are either spatially  separated (split soliton)
or located at the same optical lattice site (overlapped soliton).
The  dependence of these solitons on the  interspecies  interaction parameter
is explicitly investigated. For repulsive interspecies interaction we show the existence of a series of  critical values at which transitions  from  an initially overlapped soliton to split solitons occur. For attractive interspecies interaction only single direct  transitions  from split to overlapped BB solitons are found. The possibility to use split solitons for  indirect measurements of scattering lengths is also suggested.
\end{abstract}

\maketitle

\section{Introduction}
Bose-Einstein condensates (BECs) are fascinating tools for
simulating different physical systems. Advanced laser technology
and its successful applications to ultracold atoms have enabled us
to engineer potentials of different geometries. A well-established
technique consists in creating a linear optical lattice (LOL) by
interfering pairs of counter propagating laser beams \cite{bloch}.
On the other hand, laser beams can also be used to vary atomic
interaction periodically in space with the help of optical
Feshbach resonances \cite{theis}. Periodically modulated atomic
interaction leads to a nonlinear optical lattice (NOL). The LOL
has been used to investigate different physical phenomena in
condensed matter physics, including Bloch oscillations
\cite{morsch,salerno}, generation of coherent atomic pulses (atom
laser) \cite{andersonP}, dynamical localization
\cite{lignier,zenesini}, Landau-Zener tunneling
\cite{jona,wimb,zenesini2} and superfluid-Mott transitions~\cite{greiner}.

Interatomic interaction in BECs gives rise to a nonlinearity which
permits localized bound states to remain stable for a long time,
due to the balance between the effects of nonlinearity and
dispersion. In the presence of a LOL, the interplay between  lattice periodicity and  interatomic interaction was shown to induce  modulation instabilities of Bloch  wavefunctions near the edge bands 
\cite{KS02}, leading to the formation of localized excitations with chemical potentials inside band gaps, the so-called gap solitons (GSs).
These excitations have been  investigated both for continuous BECs, in one-dimensional \cite{alfimov,ms1,sekh1,carusotto,sekhpr} and multi-dimensional \cite{baizakov02,baizakov03,abdullaev05} settings, and for BEC arrays \cite{trombettoni,abdullaev01} in the presence of attractive and repulsive interactions. NOL can also
support special kinds of solitons both in 1D \cite{boris} and in multi-dimensional settings in combination with LOL \cite{gammal,luz}. NOLs have been used  to avoid dynamical instabilities of gap-solitons and to induce long-lived
Bloch oscillations \cite{salerno2}, Rabi oscillations \cite{bludov} and dynamical localization \cite{dyn-loc} in the nonlinear regime. For comprehensive reviews  on single-component BECs in linear and/or nonlinear optical lattices see \cite{boris,konotop,morsch1,bdz}.

On the other hand, the analysis of the physical properties of
binary mixtures of condensates still displays open issues, and represents
an interesting research topic \cite{kostov,cruz,binbec,walls,jisha,indekeu,two}.
In the past years some work has been done on the stability and
dynamics of binary BEC mixtures with both
components loaded in LOLs \cite{AdikhariMalomed} or in
NOLs \cite{abdullaev} or combinations thereof \cite{sekh2,cheng,balaz1,balaz2}.

However, BEC mixtures with one
component loaded in a LOL and the other loaded in a NOL
have not been investigated, to the best of our knowledge. This
setting is particularly interesting because it may support new
types of matter waves, due to the interplay between the different
types of OL  and the intrinsic nonlinearities. In particular, in absence of any
interaction (e.g.\ with all
scattering lengths tuned to zero), the spectrum of the component
in the LOL displays a band structure, while that of the other
component has free-particle features. It is known that for
attractive intraspecies interactions, uncoupled mixtures will
feature localized states. In this situation one can expect that a
rich variety of bound states can be formed once the interspecies
interaction is switched on.

The aim of the present paper is to study localized matter waves of  binary  BEC mixtures with one component loaded in a LOL and the other in a NOL.
In particular, we concentrate on localized states which have  chemical potentials of both components in the lower semiinfinite part of the spectrum. We call these states  bright-bright (BB)  solitons, or also  ``fundamental"  solitons, because when intraspecies scattering lengths are both negatives (the case investigated in this paper) they coincide with the ground state of the system.
We show that BB solitons can be  classified according to the distance between the lattice sites where centers of their components densities are located. Denoting these distances by $n L,\; n=0,1,2,...$, with $L$ the spatial period of the lattices (assumed to be the same for both LOL and NOL), the   $n=0$ and $n\ne0$ families are referred to as overlapped and split BB solitons, respectively. The existence and stability of these solitons are investigated both by a variational approach (VA) for the mean-field two-component Gross-Pitaevskii  equation (GPE), and by direct numerical integrations of the system. In particular, the  dependence of the existence ranges of BB soliton pairs on the  interspecies  interaction parameter, $\gamma_{12}$, is investigated. As an interesting result, we find that one can pass from one soliton family to another  by simply changing the strength of the interspecies interaction. In particular, starting from an overlapped ($n=0$) BB soliton one finds a series of  repulsive  critical values of $\gamma_{12}$ at which the transition  from  the $n$- to the $n + 1$-split BB soliton occurs as  $\gamma_{12}$ is adiabatically increased away from the uncoupling  limit ($\gamma_{12}=0$). On the contrary, for attractive interspecies interaction only direct  transition  from split  to overlapped BB solitons are possible. Since critical values at which transitions occur depend on  physical parameters of the mixture, these phenomena  suggest that split BB solitons  could be used for  indirect measurements of scattering lengths in real experiments.

The paper is organized as follows. In Section II, we introduce
the mean field equations for the coupled system and envisage a
variational study for stationary localized states. We examine the
linear stability of these states for attractive and repulsive
intercomponent interaction. In Section III, we introduce a
time-dependent variational approach, with Gaussian trial
solutions, to study different classes of BB soliton
pairs. The stability of split and overlapped families of soliton pairs
is checked by numerical integration of the mean-field equations.
In Section IV, a  numerical routine is employed to
understand the role of interspecies interaction in the splitting
mechanism, for both attraction and repulsion between different
species. Finally, in Section V we make concluding remarks.

\section{Analytical formulation}

Throughout this paper, we shall consider a
quasi-one-dimensional binary mixture of BECs, in which the
transverse motion is frozen into the ground state of a tight
transverse trapping potential, with trapping frequency
$\omega_{\perp}$. The mean-field dynamics of a mixture in which
the two species' particles have equal mass $m$ is modeled by the
coupled GPEs~\cite{stringari}
\begin{eqnarray}
\ii \hbar \frac{\partial \Psi_j}{\partial \tau} & = & \biggl(
-\frac{\hbar^2}{2m} \frac{\partial^2}{\partial s^2} +
\mathcal{V}_j(s) + 2 a_j(s) \hbar\omega_{\perp} |\Psi_j|^2
\nonumber
\\ & & \quad + 2 a_{12} \hbar \omega_{\perp} |\Psi_{3-j}|^2 \biggr)
\Psi_j,
\end{eqnarray}
where $j=1,2$ is the species index, $\mathcal{V}_j$'s are the
external trapping potentials, $a_j$'s the intraspecies scattering
lengths (which generally depend on position) and $a_{12}$ the
interspecies scattering length. The wave functions are normalized
to the numbers of particles
\begin{equation}
\mathcal{N}_j=\int ds |\Psi_j|^2.
\end{equation}

Since our system is subject to an external
potential proportional to $\cos(2 k_L s)$ generated by two counterpropagating
laser beams, the inverse wavenumber $k_L^{-1}$ and the recoil
energy $E_r=(\hbar k_L)^2 /2m$ provide natural units for length,
energy and time \cite{konotop}. To simplify the notation, we introduce the
adimensional quantities
\begin{align}\label{adimensional}
& x:=k_L s, \quad t := \frac{2 E_r}{\hbar} \tau, \nonumber \\
& V_j := \frac{\mathcal{V}_j}{2 E_r}, \quad \psi_j := \sqrt{
\frac{\hbar \omega_{\perp}}{2 E_r k_L} } \Psi_j,
\nonumber\\
& N_j := \frac{\hbar \omega_{\perp}}{2 E_r}\mathcal{N}_j, \quad \tilde{\gamma}_j:=2 a_j k_L, \quad \gamma_{12} := 2 a_{12} k_L,
\end{align}
yielding the GPEs
\begin{equation}\label{gp}
i \frac{\partial \psi_j}{\partial t} = \!\left(\! -\frac{1}{2}
\frac{\partial^2}{\partial x^2} + V_j(x) + \tilde{\gamma}_j (x)
|\psi_j|^2 + \gamma_{12} |\psi_{3-j}|^2 \!\right)\! \psi_j,
\end{equation}
and the constraint
\begin{equation}
\label{normalize}
\int dx\, |\psi_j(x)|^2 =
N_j.
\end{equation}
(Note how the variables $N_j$, which will be used throughout this article, coincide with the actual numbers of particles
$\mathcal{N}_j$ only up to a factor.) In the physical case of
interest, only the first species is subject to an external lattice
potential:
\begin{equation}\label{potentials}
V_1(x) = V_{01} \cos (2x), \qquad V_2(x)=0.
\end{equation}
The loose longitudinal harmonic trapping will be neglected, since
we will focus on states that are localized over a few lattice
sites. As for the interspecies coupling constants, we shall assume
that the interaction of the second-species particles depends on the
lattice modulation:
\begin{equation}
\label{couplings}
\tilde{\gamma}_1(x) = \gamma_1, \qquad \tilde{\gamma}_2(x) =
\gamma_2 + V_{02} \cos(2x).
\end{equation}
To allow the existence of bright solitons, which are stationary
states with both species localized, the average coupling constants
$\gamma_j$ are assumed to be negative (so that $|\gamma_2 | > |V_{02} |$).  Matter-wave bright
solitons have also been observed experimentally in trapped systems
\cite{Khaykovich}. Since the presence of the linear and nonlinear
lattice potentials in (\ref{gp}) is an obstruction to finding
exact bright soliton solutions, our study will be based on a
reasonable variational approach, with a subsequent numerical test.

\subsection{Stationary solutions: overlapped solitons}

We are interested in the stationary solutions of the coupled GPEs~(\ref{gp}). The form $\psi_j(x,t)=\phi_j(x) \e^{-\ii\mu_j t}$
yields the stationary GPEs
\begin{equation}\label{gpstat}
\!\left(\! -\frac{1}{2} \frac{\partial^2}{\partial x^2} + V_j(x)
-\mu_j + \tilde{\gamma_j}(x) |\phi_j|^2 + \gamma_{12}
|\phi_{3-j}|^2 \!\right)\! \phi_j = 0,
\end{equation}
where the external potentials and coupling constants are
given, respectively, by (\ref{potentials}) and
(\ref{couplings}). Each stationary state is characterized by the
chemical potentials $(\mu_1,\mu_2)$, which are fixed by the
normalization conditions.

\begin{figure} \centerline{
\includegraphics[width=4.3cm]{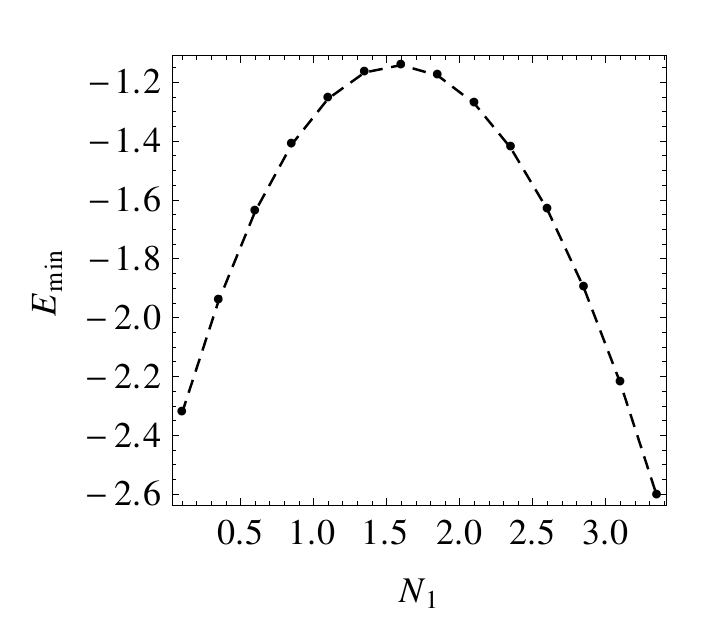}
\hskip -0.25cm
\includegraphics[width=4.3cm]{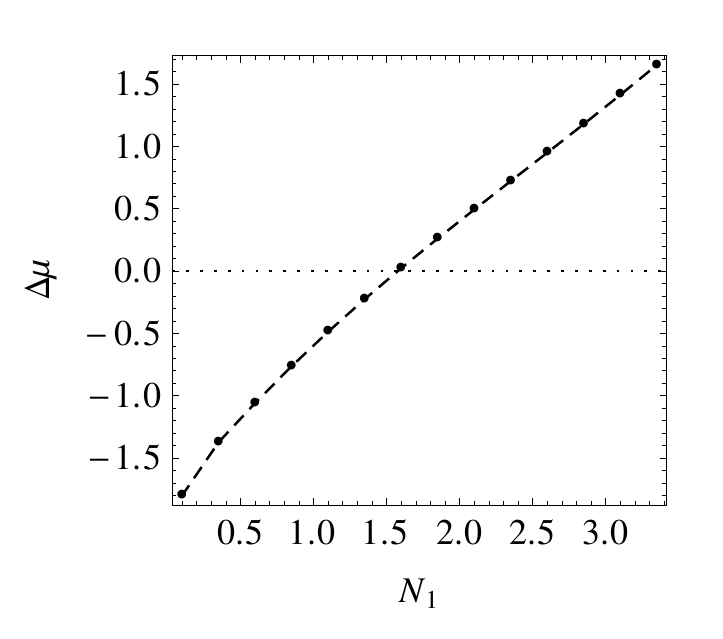}}
\caption{Left panel: dependence of the minimum energy of the
system, within the Ansatz (\ref{eq5}), vs the number of atoms $N_1$,
with $N_1+N_2=3$. Right panel: difference between chemical
potentials $\Delta_{\mu}=\mu_1-\mu_2$ vs $N_1$. The
parameters are fixed to $\gamma_1=\gamma_2=-1, V_{01}=-0.5$ and $
V_{02} = -0.25$. Energies and chemical potentials
are in units of $2E_r$. } \label{fig1}
\end{figure}

Let us assume that the intraspecies interactions are attractive
($\gamma_j<0$). Moreover, we shall focus on the
case $V_{01},V_{02}<0$: in this situation, due to the (linear and
nonlinear) trapping mechanisms, density profiles peaked around the
points where $\cos (2x)=1$ are energetically favorable for both
species. In order to investigate the features of BB soliton
pairs, we choose a Gaussian trial solution
\begin{equation}\label{eq5}
\phi_{j}(x)=A_j \exp\left[-{x^2}/{2 a_j^2}\right].
\end{equation}
Since the amplitudes $A_j$ and the widths $a_j$ are bound by the
normalization conditions
\begin{equation}
N_j= \int \d x\, |\phi_j(x)|^2 = \sqrt{\pi} a_j A_j^2,
\end{equation}
the functions (\ref{eq5}) have only one free parameter. Moreover,
this class of trial solutions fits
\emph{overlapped} BB solitons, with the peak  of their
densities sitting at the same position, say, at $x=0$. Since, due to attractive
interspecies interactions, the superposition of densities lowers
the energy of the system, we expect the most energetically
favorable soliton pair to be overlapped.

At fixed numbers of particles, the Gross-Pitaevskii energy
functional for $\phi_j$ in the class (\ref{eq5}) can be viewed as
a function of the soliton width:
\begin{eqnarray}\label{eq6}
E & = &\int\! \d x \Biggl[ \frac{1}{2} \sum_{j=1,2} \Biggl( \left| \frac{\partial
\phi_j}{\partial x}\right|^2 + \tilde{\gamma}_j(x) |\phi_j|^4 \Biggr) \nonumber \\ &  & \qquad\quad +
\gamma_{12} |\phi_1|^2 |\phi_2|^2 + V_1(x) |\phi_1|^2 \Biggr]
\nonumber
\\ & = & \frac{1}{\sqrt{8\pi}}
\left(\frac{\sqrt{\pi}N_1}{\sqrt{2}a_1^2} +
\frac{\sqrt{\pi}N_2}{\sqrt{2}a_2^2} +\sqrt{8\pi } V_{01} N_1
\e^{-a_1^2} \right.\nonumber\\&+&\left.\frac{V_{02} N_2^2}{a_2}
\e^{-\frac{a_2^2}{2}} + \frac{{\gamma_1}N_1^2}{a_1} +
\frac{\gamma_2N_2^2}{a_2} + \frac{ {\gamma_{12}}
N_1N_2}{\sqrt{a_1^2+a_2^2}}\right). \nonumber \\
\end{eqnarray}
The optimal width values $(\bar{a}_1(N_1,N_2),\bar{a}_2(N_1,N_2))$ are determined by
\begin{equation}
\left.\frac{\partial E}{\partial a_j} \right|_{a_k=\bar{a}_k} = 0
\qquad \text{for } j=1,2,
\end{equation}
and fix the trial ground state. The corresponding energy will be
denoted by
\begin{equation}
E_{\min} (N_1,N_2) = E|_{a_j=\bar{a}_j (N_1,N_2)}.
\end{equation}
The chemical potentials $\mu_j = \partial E_{\min} / \partial N_j$ can be used
to test the linear stability of the ground state solution through
the Vakhitov-Kolokolov criterion \cite{vakhitov}.


\begin{figure}
\vskip -.75cm
\centerline{
\includegraphics[width=4.5cm]{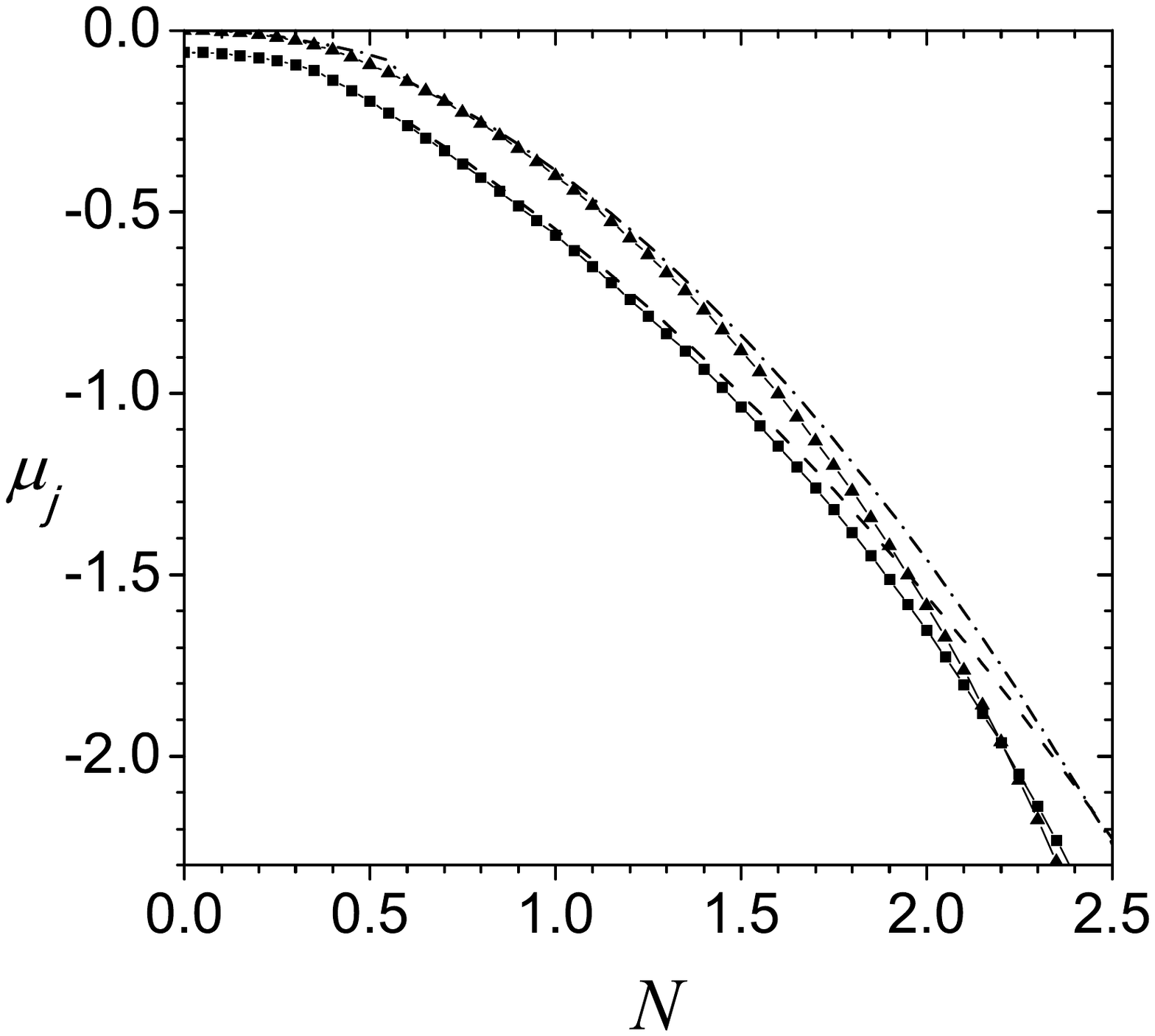}
\hskip -0.5cm
\includegraphics[width=4.5cm]{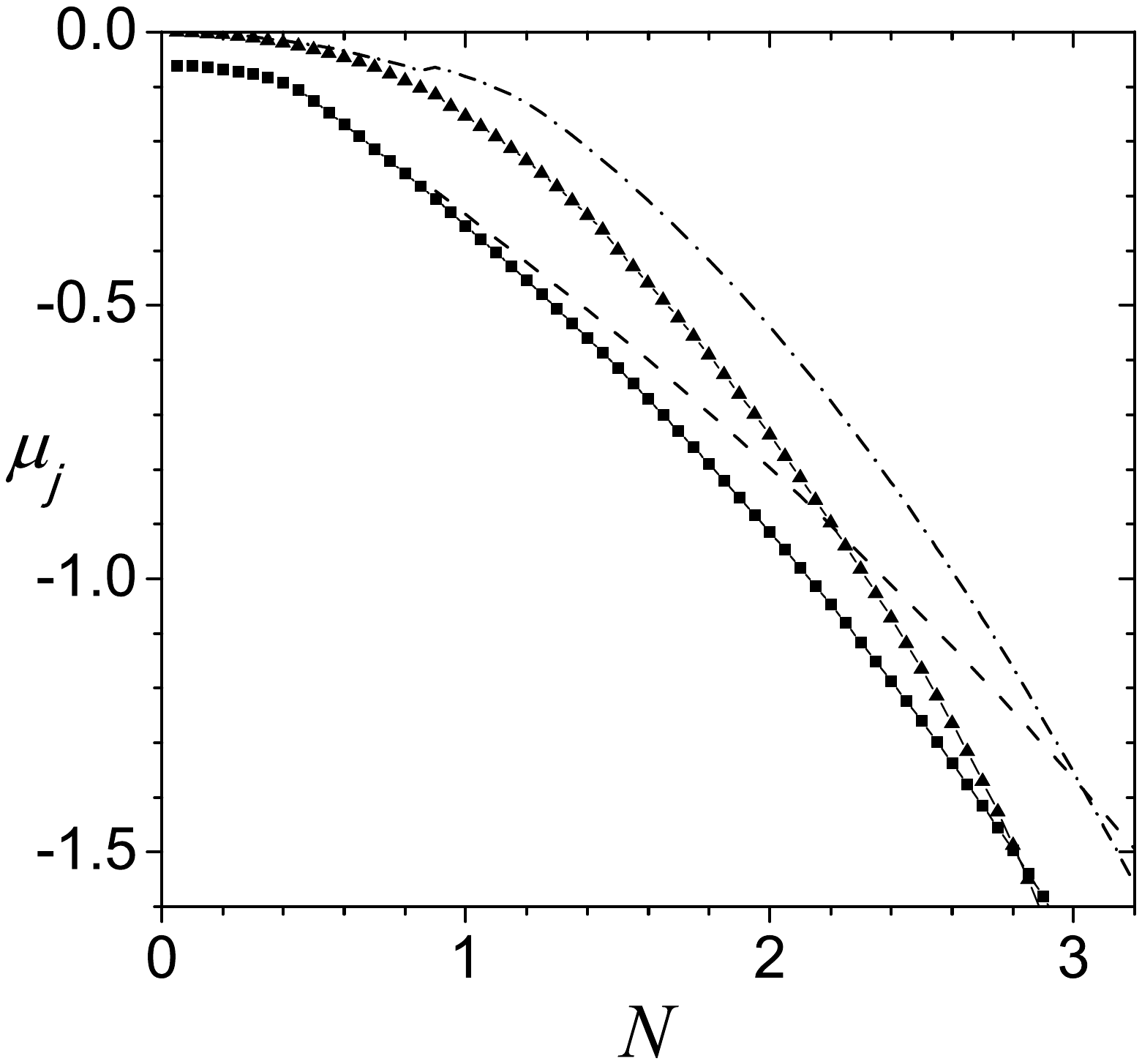}
}
\vskip -2.1cm
\caption{Existence curves in the
$\mu_i-N$ plane ($N \equiv N_1=N_2$) for BB solitons of the GPEs with
attractive ($\gamma_{12}=-0.5$, left) and repulsive
($\gamma_{12}=0.1$, right) interspecies interaction. Other
parameters are fixed as $\gamma_1=\gamma_2=-1,  V_{01}=-0.5, V_{02}
=-0.25$. Square and triangle lines denote numerical results  obtained
from Eqs.\ (\ref{gp}) by imaginary time relaxation method
and refer to
the first and the second component,
respectively. Dashed and dot-dashed lines represent the corresponding results
obtained from the
minimization of the energy functional. }\label{fig1ms}
\end{figure}

Relevant properties of the overlapped solitons can be inferred
from the energy functional in Eq.~(\ref{eq6}) and the chemical
potential. If one keeps constant the total number of atoms $N=N_1+N_2$, the change in the energy $E_{\min}(N_1,N_2)$ of the
trial ground state with $N_1$ (or equivalently $N_2$) can be
analyzed. Let us fix for definiteness
$\gamma_1=\gamma_2=-1$, $\gamma_{12}=-0.5$, $V_{01}=-0.5$,
$V_{02}=-0.25$. Throughout the paper, numbers $N_j$ of order one will be extensively used: recall that they are related to the actual numbers of atoms $\mathcal{N}_j$ by the factor $2E_r/\hbar\omega_{\perp}$ thru Eq.\ (\ref{adimensional}). In an experiment with ${}^7$Li atoms ($m\simeq 1.2\times 10^{-27}$ kg), with a transverse trapping potential $\omega_{\perp}\simeq 2\pi\times 700$ Hz \cite{Khaykovich,cornish} and a laser wave number $k_L=5\times 10^{-7}\, \mathrm{m}^{-1}$, the conversion factor is of the order of $10^4$.

In Fig.~\ref{fig1} we set the (scaled) number of
particles to $N=3$, and study the behavior of the minimal energy
and the difference in chemical potentials of the soliton pair. The
evident asymmetry in the plot of the energy $E(N_1,N-N_1)$ (left
panel) with respect to $N_1=N/2$ is related to the inhomogeneity
of the lattice potentials and the self-interactions for the two
components. Moreover, the difference in chemical potentials
$\Delta\mu=\mu_1-\mu_2$ (right panel) can lead to changes in the
numbers of particles if the system is in contact with a particle
reservoir or if a transition mechanism between the two species is
present.

Existence curves in the $\mu-N$ plane are plotted
in Fig.~\ref{fig1ms} for the case of equal attractive
intraspecies interactions and for both attractive (left panel)
and repulsive (right panel) interspecies interaction. To reduce
the number of parameters we have considered the case
$N_1=N_2\equiv N$. The dotted lines refer to the VA results
obtained from the numerical minimization of the energy
(\ref{eq6}), while the solid lines represent the corresponding
curves obtained from numerical relaxation method and
self-consistent diagonalization of the stationary GPEs in
Eq.~(\ref{gpstat}) \cite{ms1,abdullaev}. From this plots one can
see that $\d \mu_j/\d N<0$, both for the VA and for the GPE curves,
which implies, according to the Vakhitov-Kolokolov criterion, that
BB solitons are linearly stable. This result is also confirmed by
direct numerical time integrations of the two component GPEs~(\ref{gp}) (not shown here).

\begin{figure}
\vskip -0.75cm \centerline{
\includegraphics[width=4.8cm]{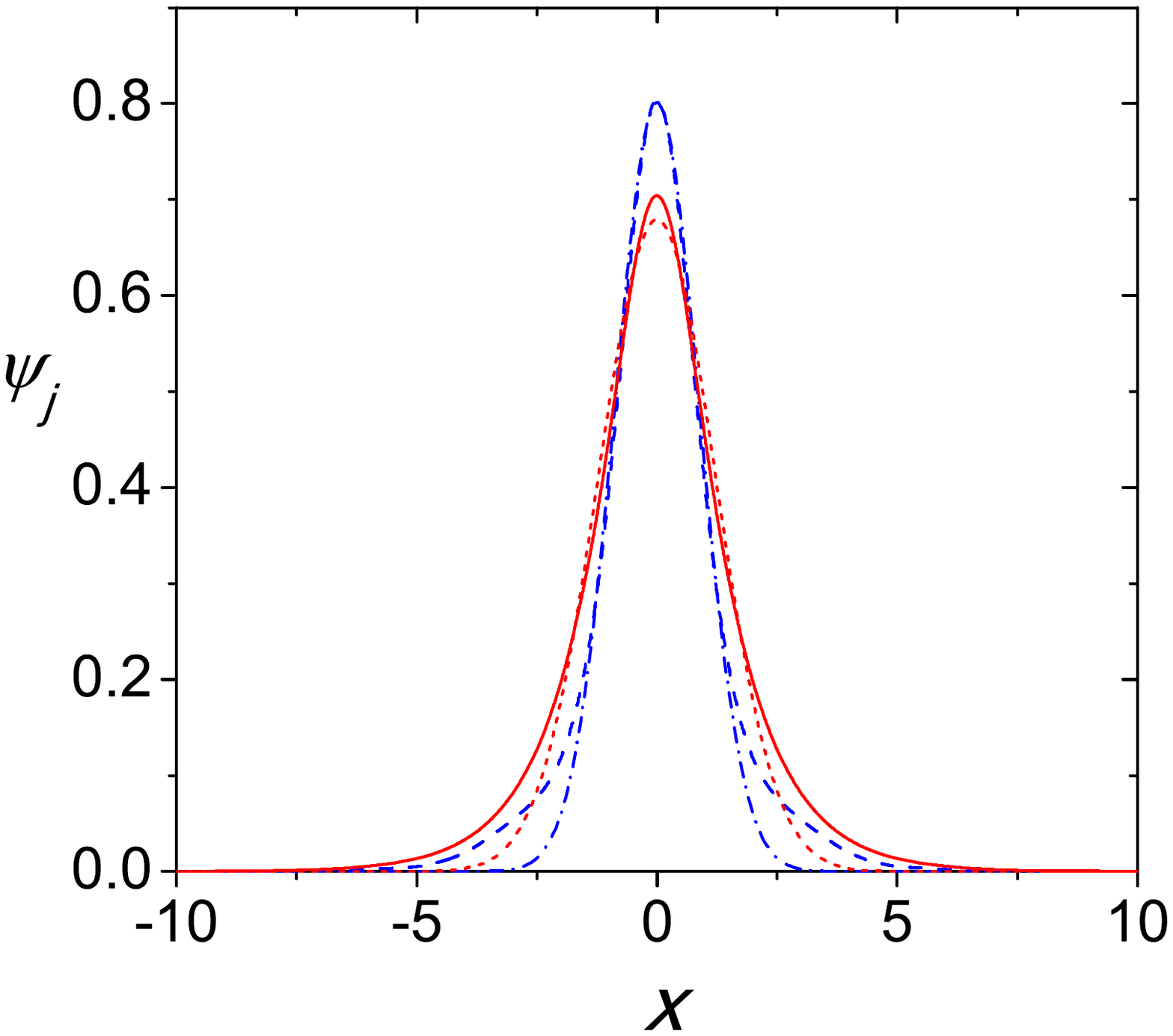}
\hskip -0.5cm
\includegraphics[width=4.8cm]{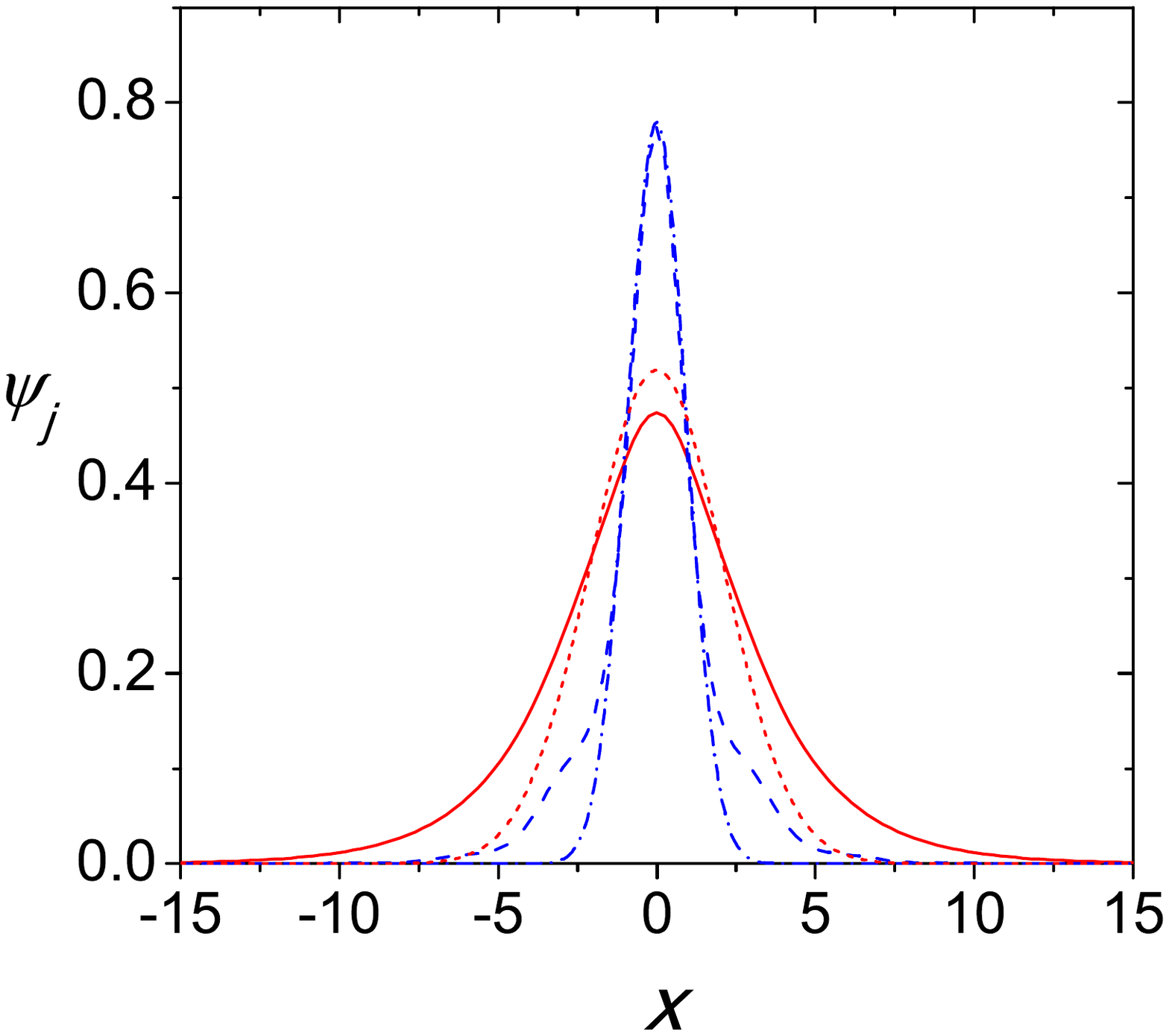}
}
\vskip -2.1cm
\caption{Profiles of BB solitons with $N_1=N_2=N$ corresponding
to the $N=1$ points on the curves
of Fig.~\ref{fig1ms} for attractive ($\gamma_{12}=-0.5$, left
panel) and repulsive ($\gamma_{12}=0.1$, right panel) interspecies
interaction.  Blue (dashed and dot-dashed) and red (continuous and dotted)
curves refer to first and second component, respectively. In both panels
continuous and dashed lines refer to numerical solutions of Eqs.~(\ref{gp})
found by relaxation method, while dotted and dot-dashed lines to
variational analysis.}
\label{fig2ms}
\end{figure}

From Fig.~\ref{fig1ms} one can see that, while in
the attractive case the agreement is quite good for a wide range
of $N$ (this is true also for relatively large values of
$\gamma_{12}$), in the repulsive case deviations of the VA and
numerical curves are larger. This discrepancy is due to the fact
that the Gaussian Ansatz becomes less accurate in the repulsive
case, in which the interspecies interaction
reduces the stability of the overlapped configuration, as one can
see from Fig.~\ref{fig2ms} where VA and GPE solitons profiles are
compared. However, the accuracy of the VA result increases as
positive $\gamma_{12}$ values are decreased towards the uncoupling
limit $\gamma_{12}=0$. Also note the existence of points where the
chemical potential curves intersect, both for attractive and
repulsive interspecies interactions. At these points, BB
solitons, having the same number of atoms and the same chemical
potentials (related to their width), will have equal VA profiles
for the two components. Despite the discrepancy in the location of
the intersection point, equality of profiles is well confirmed by
numerical GPE results.

\section{Split BB solitons and dynamical properties}

In the previous section, the choice of Gaussian
trial wave functions (\ref{eq5}) aimed at studying overlapped
BB soliton configurations, with the peaks of the two
density profiles coinciding at $x=0$. Due to attractive
interspecies interaction, this configuration is expected to be the
lowest-energy BB soliton pair. It is possible however to extend
the analysis to \textit{split BB solitons}, in which the centers
of mass of the two species do not coincide.

Let us initially consider the uncoupled limit
$\gamma_{12}=0$. In the case $V_{01}<0$ and $V_{02}<0$, one
expects an infinite set of degenerate energy-minimizing BB
solitons, since the centers of mass of each species can be located
at any point $x_{0j}$ such that $\cos(2x_{0j})=0$, regardless of
the other species' density profile. These minimizing
configurations can be classified in families
\begin{equation}
\label{BBNfamilies}
BB_n(\gamma_{12}=0), \qquad
\textrm{with} \; n\in\mathbb{N},
\end{equation}
according to the absolute distance between
the centers of mass:
\begin{equation}\label{deltax0}
\Delta x := |x_{02}-x_{01}| \in \left(
\!\left(n-\frac{1}{2}\right)\! \pi, \!\left(n+\frac{1}{2}\right)\!
\pi \right).
\end{equation}
Clearly, the energy of the BB soliton configurations is the
same for all families, $E_n(\gamma_{12}=0)=E_0$. When the effect
of the interspecies coupling $\gamma_{12}$ can be treated as a
small perturbation, one expects the existence of stationary
Gross-Pitaevskii solutions close to the ones at $\gamma_{12}=0$,
which can  still be classified in families $BB_n(\gamma_{12})$
according to the criterion (\ref{deltax0}). However, if
$\gamma_{12}<0$, the interspecies attraction will break the
energetic degeneracy in favor of the overlapped configuration
$BB_0$. Thus, the split solitons in $BB_{n>0}(\gamma_{12})$ become
metastable. While configurations with a very large distance
between the two species are almost unaffected by interspecies
interactions, larger values of $|\gamma_{12}|$ weaken the (local)
stability of solitons with small $n$, since attractive
interactions can give a sufficient amount of energy to overcome
the (linear or nonlinear) potential barrier and reduce the
distance between the centers of mass. Thus, one expects a critical value $\gamma_{\mathrm{cr}}^{(n)}$, such that for
$\gamma_{12}<\gamma_{\mathrm{cr}}^{(n)}$ metastable solutions in
$BB_n$ would no longer exist.

\begin{figure}
\centerline{
\includegraphics[width=4.2cm]{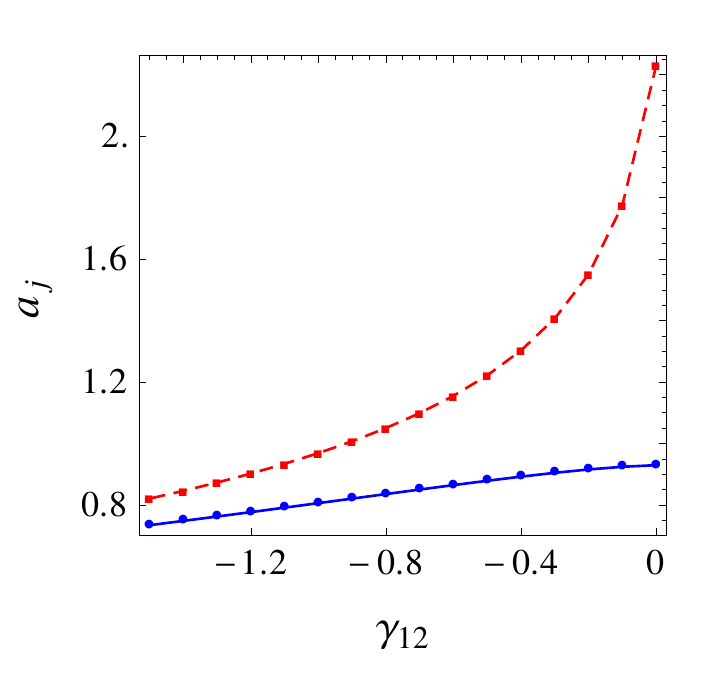}
\includegraphics[width=4.2cm]{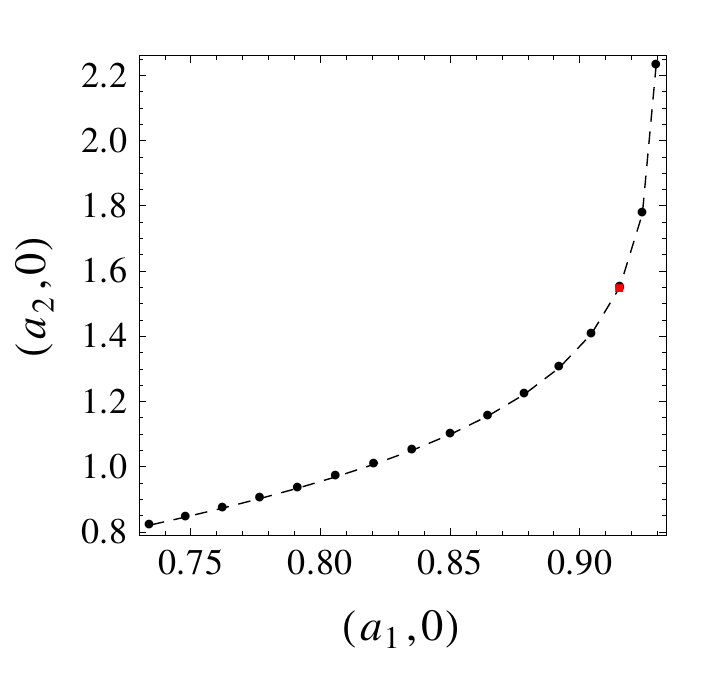}}
\centerline{
\includegraphics[width=4.2cm]{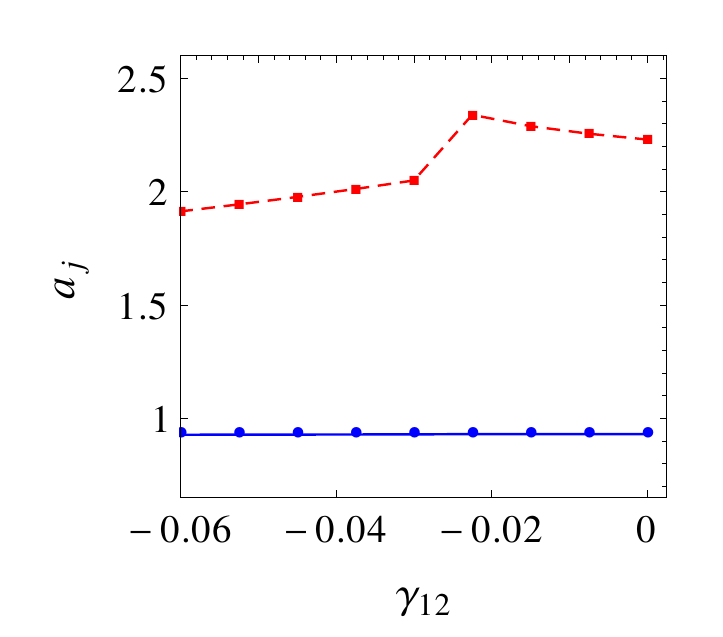}
\includegraphics[width=4.2cm]{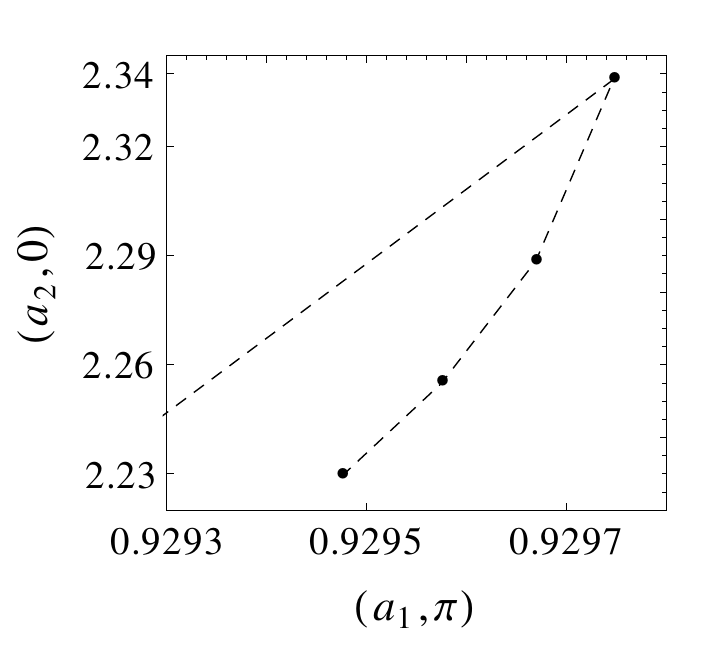}}
\centerline{
\includegraphics[width=4.2cm]{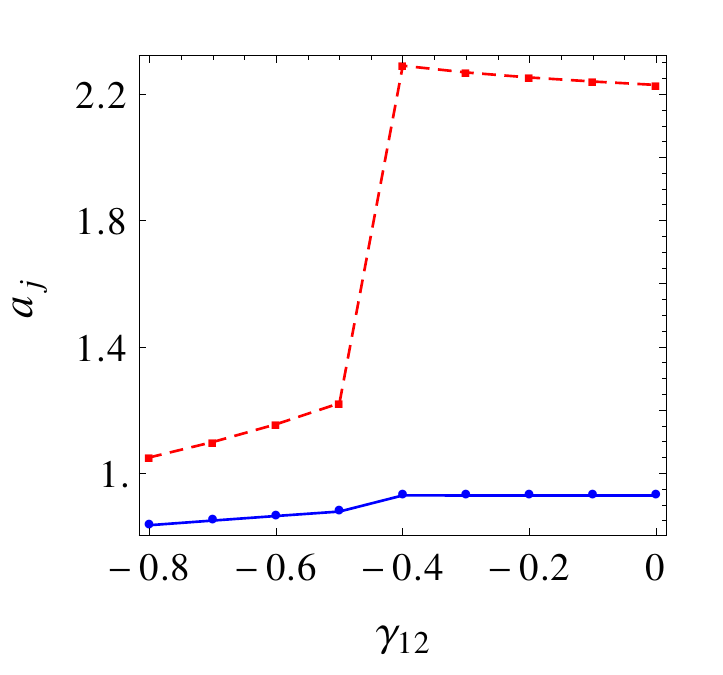}
\includegraphics[width=4.2cm]{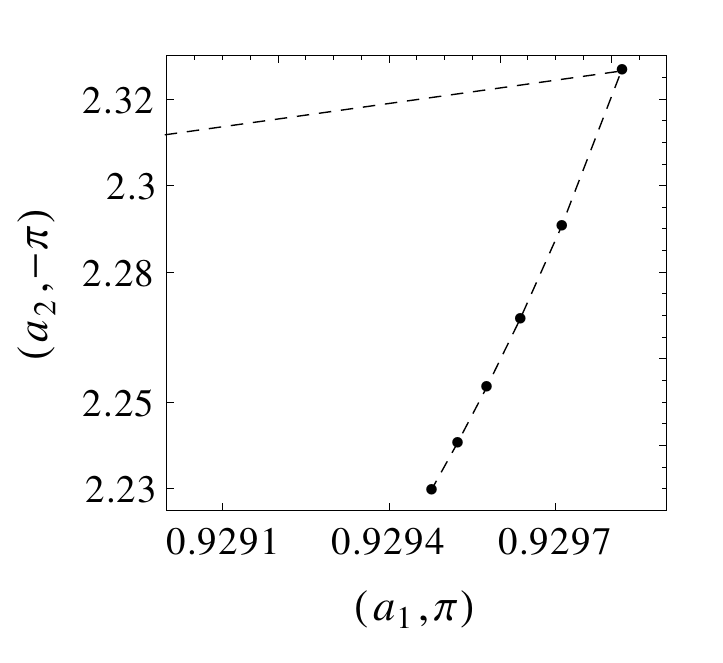}}
\caption{Numerical minimization of the
pseudopotential $\Pi$ [see (\ref{eq15})-(\ref{eq16}) with soliton
peaks around $(x_{01},x_{02})=(0,0)$ (top line),
$(x_{01},x_{02})=(\pi,0)$ (central line),
$(x_{01},x_{02})=(\pi,-\pi)$ (bottom line). Plots in the left
column represent the variation of the optimal soliton widths $a_1$
(solid blue lines) and $a_2$ (dashed red lines) with
$\gamma_{12}$. Plots in the right column show the behavior of one
optimal width with respect to the other. The jumps observed in the
central and bottom lines (right column) are due to the
disappearance of local minima with $\Delta x\simeq\pi$ and
$\Delta x \simeq 2\pi$ as $\gamma_{12}$ decreases under a critical
value. Minima found below the critical $\gamma_{12}$ coincide
with the values in the top plots and are thus not shown in the
right column. }\label{fig2}
\end{figure}

\begin{center}
\begin{table}
\begin{tabular}{|c|c|c|c|c|c|}
\hline
$\gamma_{12}$&$ n $&$ \quad \Delta x \quad $& $\quad a_1 \quad$ &$\quad a_2 \quad $&$\quad \Pi_{\min} \quad$\\
\hline
0 & 0,1,2 & $0,\pi,2\pi$ & $0.929$ & $2.229$ & $-0.177$ \\
\hline
$-0.25$ & 0 & 0 & $0.910$ & $1.472$ & $-0.249$\\
\hline
$-0.25$ & 1 & \multicolumn{4}{c}{no local minimum} \vline \\
\hline
$-0.25$ & 2 & $2.991\,\pi$ & $0.930$ & $2.260$ & $-0.177$ \\
\hline
\end{tabular}
\caption{Locally stable BB soliton pairs for $N_1=1$, $N_2=1$, $\gamma_1=-1$,
$\gamma_2=-1$, $V_{01}=-0.5$, $V_{02}=-0.25$, belonging to the classes $BB_n$ with $n=0,1,2$. The optimal parameters $\Delta x$, $a_1$ and $a_2$ (units of $k_L^{-1}$) are obtained by minimizing the global potential $\Pi$ (units of $2E_r$).}\label{tab1}
\end{table}
\end{center}

In the following, we will numerically analyze the
existence of split BB soliton pairs, as well as their energetic and dynamical behavior. To this end, we shall generalize our Ansatz to
include the positions of the component density centers as free
parameters. We will also consider time-dependent parameters to
investigate the dynamics of the system. The Gross-Pitaevskii
equations (\ref{gp}) can be restated as a variational problem
\cite{anderson}
\begin{eqnarray} \label{eq8}
\delta \int {\cal{L}}\left(\psi_j,\psi^*_j,\frac{\partial
\psi_j}{\partial x},\frac{\partial \psi^*_j}{\partial
x},\frac{\partial \psi_j}{\partial t},\frac{\partial
\psi^*_j}{\partial t}\right)\! \d x \,\d t=0,
\end{eqnarray}
where the Lagrangian density reads
\begin{eqnarray}\label{eq9}
{\mathcal L} & = & \frac{1}{2} \sum_{j=1}^{2} \left[ i \!\left(
\psi^*_j \frac{\partial \psi_{j}}{\partial t} - \psi_j
\frac{\partial \psi^* _{j}}{\partial t}\right)\! - \!\left|
\frac{\partial \psi_{j}}{\partial x} \right|^2 - \gamma_j
|\psi_j|^4 \right] \nonumber \\ &  & - V_{01} \cos (2 x) |\psi_1|^2
- \frac{V_{02}}{2} \cos (2 x) |\psi_2|^4 \nonumber \\ &  & -
\gamma_{12} |\psi_1|^2 |\psi_2|^2 .
\end{eqnarray}
We generalize the Gaussian Ansatz (\ref{eq5}) to
\begin{eqnarray}\label{eq10}
\psi_{j}(x,t)&=& A_j \exp\left[\ii\frac{\d{x}_{0j}}{\d t}
(x-x_{0j}) + \ii \phi_j(x-x_{0j})^2+\ii \theta_j\right]
\nonumber\\
& &\times
\exp\left[-\frac{(x-x_{0j})^2}{2
a_j^2}\right],
\end{eqnarray}
where the variational parameters
$(A_j,a_j,x_{0j},\phi_j,\theta_j)$ for $j=1,2$ are generally
time-dependent and represent, respectively, the amplitude, width,
center-of-mass position, frequency chirp and overall phase of a
soliton in the $j$-th component. The Lagrangian $L$ for the trial
wave functions (\ref{eq10}), obtained by integrating the
Lagrangian density in (\ref{eq9}), reads
\begin{eqnarray} \label{eq11}
L & =& \int_{-\infty}^{+\infty} \,\,{\cal L}\,\,\d x\nonumber\\
&=& \frac{\sqrt{\pi}}{4} \left(\sum_{j=1}^{2}
\left[\frac{A_j^2}{a_j} +\sqrt{2}
\gamma_j a_j A_j^4+4 a_j^3 A_j^2 \phi_j^2 \right.\right. \nonumber \\
& & - \left.\left. 2 a_j A_j^2
\left(\frac{\d x_{0j}}{\d t}\right)^2 + 4a_j A_j^2
\frac{{\d\theta}_j}{\d t} + 2 a_j^3 A_j^2 \frac{\d \phi_j}{\d t}
\right] \right.\nonumber\\
& & + \left.4 V_{01} \e^{-a_1^2} a_1 A_1^2
\cos(2 x_{01}) \right.\nonumber\\
& & + \left.\sqrt{2} V_{02}
\e^{-\frac{1}{2} a_2^2} a_2 A_2^4 \cos(2 x_{02}) \right. \nonumber
\\
&  & + \left. \frac{4\gamma_{12} \,
\e^{-\frac{(x_{01}-x_{02})^2}{{a_1}^2+{a_2}^2}} a_1 A_1^2 a_2
A_2^2}{\sqrt{{a_1}^2+{a_2}^2}}\right).
\end{eqnarray}
Note the dependence of the interaction term on $\Delta x$, defined
in (\ref{deltax0}).

\begin{figure}
\centerline{
\includegraphics[width=4.5cm]{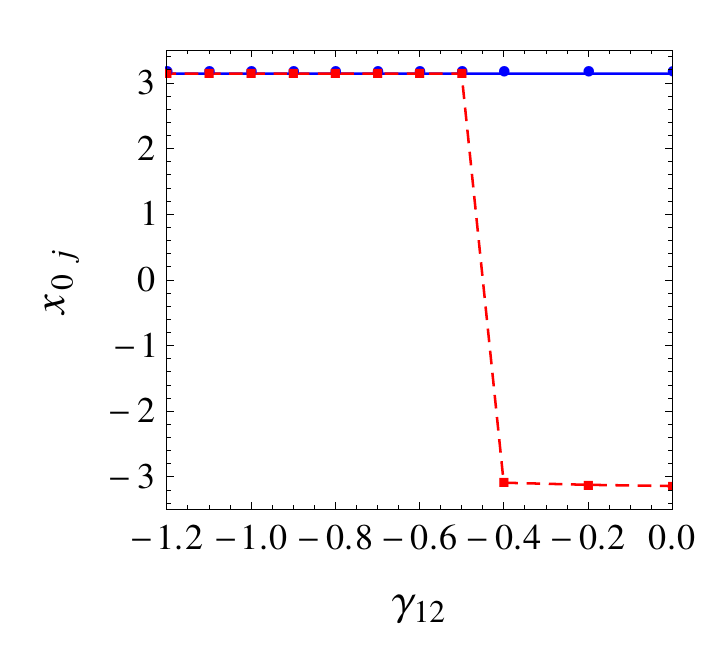}
\includegraphics[width=4.5cm]{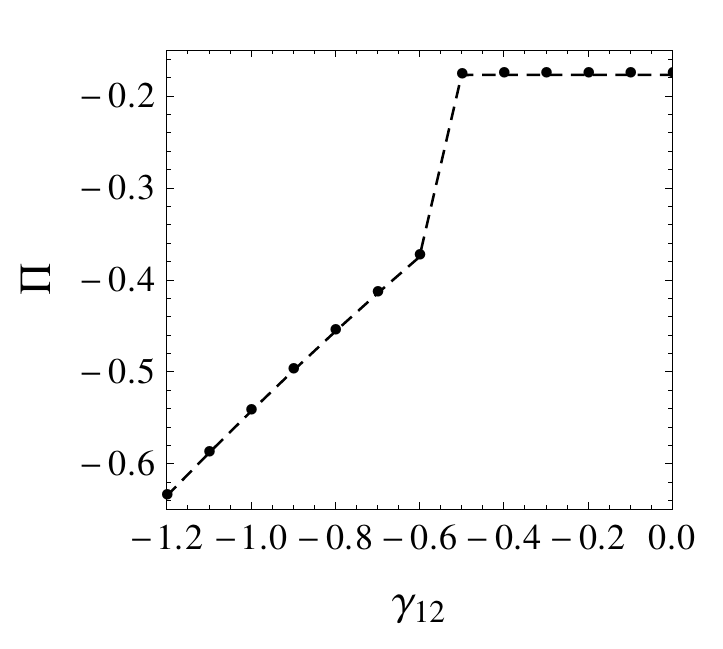}}
\caption{Left panel. Peak positions $x_{01}$ (solid blue line) and
$(x_{02})$ (dotted red line) of optimal BB profiles, obtained by
searching the local minimum around $(x_{01},x_{02})=(\pi,-\pi)$,
for different values of $\gamma_{12}$. The sudden jump towards
$(\pi,\pi)$ is due to the disappearance of the local minimum (see also
Fig.~\ref{fig2}, bottom panels). Right panel. Local minimum value of the effective
potential energy in Eq.\ (\ref{eq15}).}
\label{fig5a}
\end{figure}

The functional derivatives of $L$ with respect to the variational
parameters yield a set of Euler-Lagrange equations. After
appropriate manipulation, we can obtain a picture for the dynamics
of the soliton pairs. In particular, the equations
\begin{eqnarray}
\frac{\d}{\d t} (\sqrt{\pi}a_jA_j^2) & = & 0, \label{eq12} \\
-2a_j \phi_j + \frac{\d{a}_j}{\d t} & = & 0 \label{eq13}
\end{eqnarray}
can be interpreted as dynamical constraints on the amplitude
(particle numbers conservation) and the frequency chirp. Taking
into account (\ref{eq12})-(\ref{eq13}), the equations of motion
for the centers of mass and the width can all be derived from the
effective potential
\begin{eqnarray}\label{eq15}
\Pi\left(a_1,a_2,x_{01},x_{02}\right) & =
&\Pi_1\left(a_1,x_{01}\right)+\Pi_2\left(a_2,x_{02}\right)
\nonumber \\ & + &\Pi_{12}\left(a_1,x_{01},a_2,x_{02}\right),
\end{eqnarray}
with
\begin{subequations}
\begin{eqnarray}\label{eq16}
\Pi_1\left(a_1,x_{01}\right)&=&\frac{N_1}{4 a_1^2}+
\frac{\gamma_1N_1^2}{2\sqrt{2 \pi } a_1}\nonumber\\
& &+ \e^{-{a_1}^2}N_1 V_{01} \cos(2 x_{01}),
\end{eqnarray}
\begin{eqnarray}\label{eq17}
\Pi_2\left(a_2,x_{02}\right)&=&\frac{N_2}{4 a_2^2}+
\frac{\gamma_2N_2^2}{2\sqrt{2 \pi } a_2}
\nonumber\\
& & + \frac{\e^{-\frac{1}{2} {a_2}^2}N_2^2 V_{02}\cos(2
x_{02})}{2 \sqrt{2 \pi } {a_2}},
\end{eqnarray}
\begin{eqnarray}\label{eq18}
\Pi_{12}\left(a_1,x_{01},a_2,x_{02}\right)=\frac{\e^{-\frac{(x_{01}-x_{02})^2}{{a_1}^2+{a_2}^2}}
\gamma_{12}N_1N_2}{\sqrt{\pi } \sqrt{{a_1}^2+{a_2}^2}},
\end{eqnarray}
\end{subequations}
through the Newton-like equations
\begin{equation}\label{eq14}
\frac{\d^2}{\d t^2} \begin{pmatrix}
N_1x_{01} \\
N_2x_{02} \\
(N_1/2) a_1 \\
(N_2/2) a_2
\end{pmatrix} =
-\begin{pmatrix}
\partial_{x_{01}} \\
\partial_{x_{02}} \\
\partial_{a_{1}} \\
\partial_{a_{2}}
\end{pmatrix}\Pi\left(a_1,a_2,x_{01},x_{02}\right).
\end{equation}
The problem of finding stationary solutions within the Ansatz
(\ref{eq10}) thus reduces to searching the minima of the effective
potential. The equations (\ref{eq14}) can also be
linearized around the stable equilibrium points for $\Pi$, to
obtain information on small center-of-mass and width oscillations.
On the other hand, due to the (still restrictive) form of the
trial wave functions, the far-from-equilibrium dynamics of
Eqs.~(\ref{eq14}) cannot be considered physically relevant (see
the discussion below.) In Table \ref{tab1}, we show some illustrative results of the optimal parameters $\Delta x$, $a_1$ and $a_2$ for solitons in the classes $BB_n$ with $n=0,1,2$. The numbers of particles are fixed to $N_1=N_2=1$. At $\gamma_{12}=0$, the three minima are degenerate. At $\gamma_{12}=-0.25$, the split solitons in the family $BB_2$ are locally stable [see Eq.\ (\ref{BBNfamilies})], while
the global minimum is, as expected, the overlapped configuration.
Instead, no local minimum can be found in the family $BB_1$, indicating that $\gamma_{\mathrm{cr}}^{(1)} > -0.25$.

A more systematic picture of the existence and behavior of
solitons for small $n$ with varying $\gamma_{12}$ is given in
Fig.~\ref{fig2}: a numerical minimization procedure is used to find stable configurations with their centers of mass close to some initial points $(\tilde{x}_{01},\tilde{x}_{02})$. A minimum with centers of mass around $(\tilde{x}_{01},\tilde{x}_{02})=(0,0)$ can be found for all negative $\gamma_{12}$ (top panels). On the other hand,
the search of minima whose centers of mass are close to $(\tilde{x}_{01},\tilde{x}_{02})=(\pi,0)$ and $(\tilde{x}_{01},\tilde{x}_{02})=(-\pi,\pi)$ yields discontinuous behaviors in the optimal parameters (central and bottom panels). The discontinuities are present because solitons in $BB_1$ (respectively $BB_2$) exist only for
$\gamma_{12}>\gamma_{\mathrm{cr}}^{(1)}$ (respectively
$\gamma_{12}>\gamma_{\mathrm{cr}}^{(2)}$), while for more
negative values the algorithm actually finds the global minimum
belonging to $BB_0$. It is also worth noticing
that in Fig.~\ref{fig2} the optimal widths $a_1$ and $a_2$ of the overlapped solitons
decrease with $|\gamma_{12}|$. In the case of split solitons, the
amplitudes remain almost constant, with a slight \textit{increase} (more evident
in $a_2$) with $|\gamma_{12}|$, due to the attraction exerted between densities.

These findings are corroborated by the behavior of the center-of-mass positions displayed in Fig.\ \ref{fig5a}, where the displacement of the center of mass of the second species with increasing interspecies interaction is observed. The situation is the same as that depicted in Fig.\ \ref{fig2}, bottom panels. In the left panel of Fig.\ \ref{fig5a}, the jump of $x_{02}$ as $|\gamma_{12}|$ is decreased signals the disappearance of the local energy minimum. The value of the local minimum of the effective potential energy (\ref{eq15}) is shown in the right panel.

\begin{figure}
\centerline{
\includegraphics[width=4.3cm]{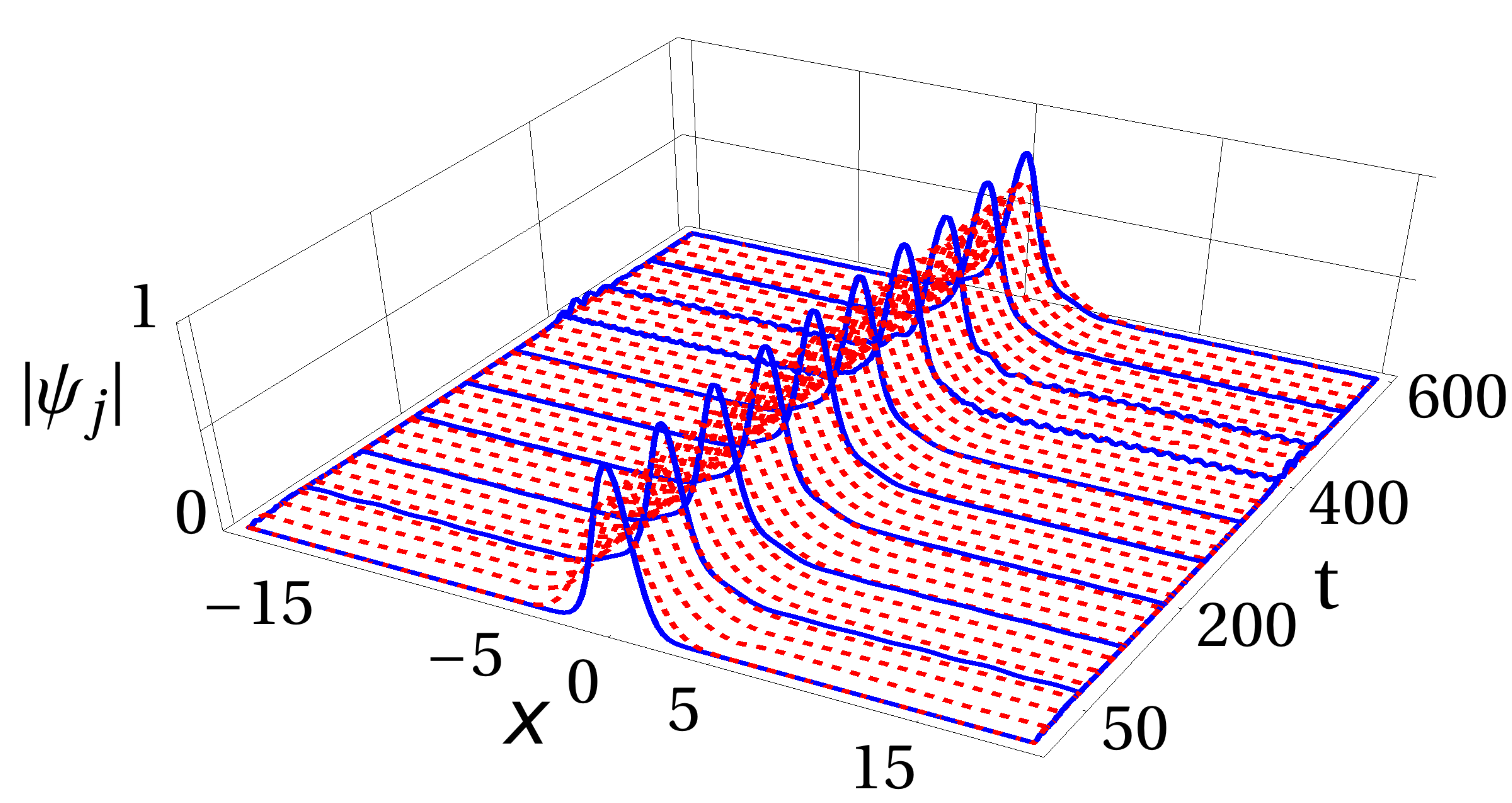}
\hskip 0.1cm
\includegraphics[width=4.3cm]{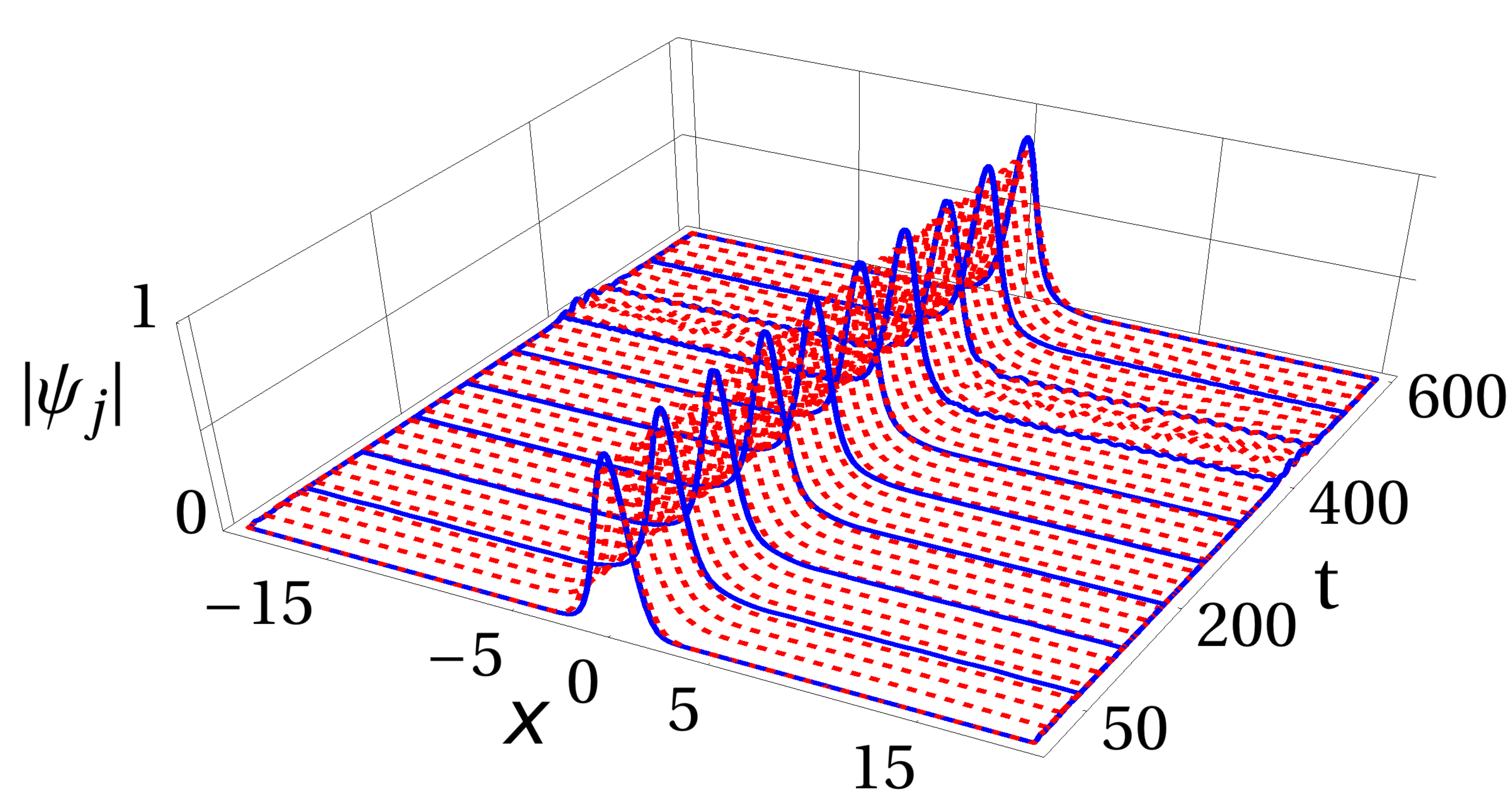}}
\caption{Time evolution of density profiles for overlapped
solitons in the first (blue solid lines) and second (red dashed
lines) species for $\gamma_{12}=-0.25$ (left panel) and
$\gamma_{12}=-1$ (right panel). The results are obtained by direct
numerical integration.}\label{fig4}
\end{figure}

\begin{figure}
\centerline{
\includegraphics[width=4.3cm]{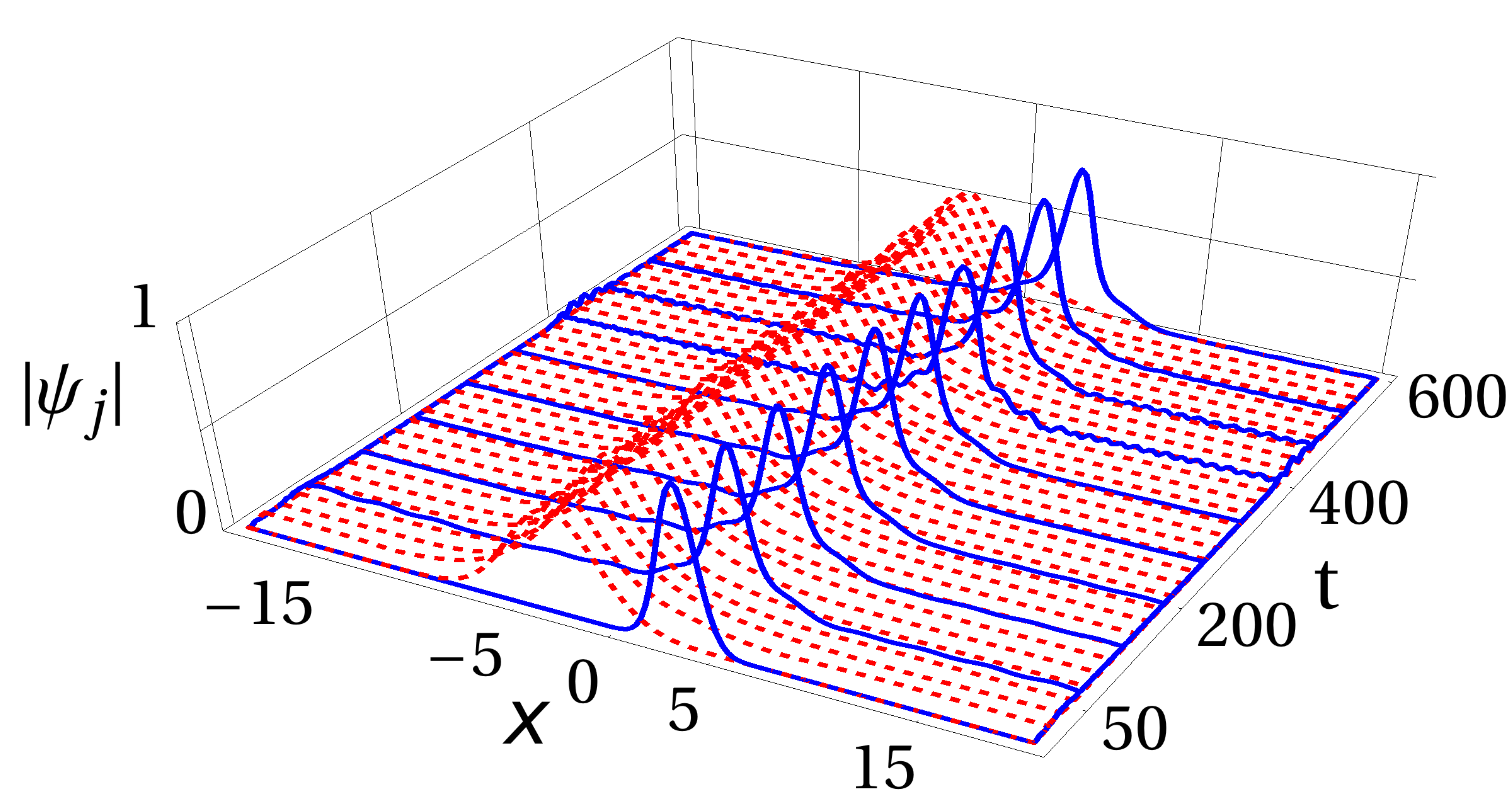}
\hskip 0.1cm
\includegraphics[width=4.3cm]{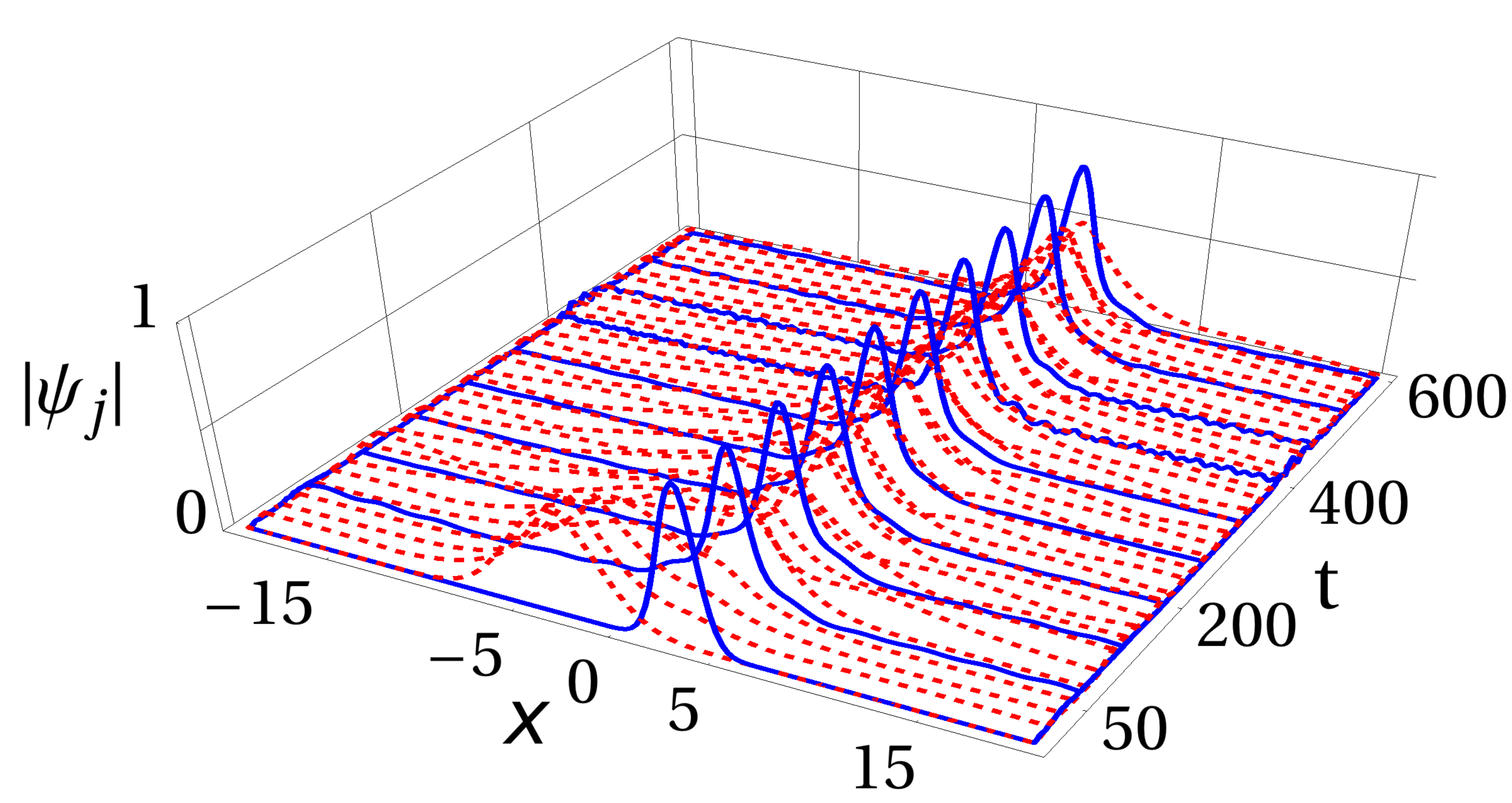}}
\caption{Time evolution of density profiles for split solitons
in the first (blue solid lines) and second (red dashed lines)
species for $\gamma_{12}=-0.25$ (left panel) and
$\gamma_{12}=-0.38$ (right panel). In the latter case, numerical
integration shows the instability of the soliton
pair.}\label{fig5}
\end{figure}

In order to check the existence and stability of BB soliton pairs
as approximate solutions of the GPEs, we employ a numerical simulation
of the dynamics generated by~(\ref{gp}). First, we have checked the
stationarity of overlapped soliton pairs, localized around $x=0$.
It is possible to verify, for different values of $\gamma_{12}$, that
the soliton pair determined by the minimization procedure is
stationary within very good approximation. In Fig.~\ref{fig4},
the time evolution of the overlapped solitons is represented for
$\gamma_{12}=-0.25$ (left) and $\gamma_{12}=-1$ (right). Then, we
have tested the behavior of split soliton pairs in $BB_2$ in
different regimes. In the case $\gamma_{12}=-0.25$, which is
larger than the critical value $\gamma_{\mathrm{cr}}^{(2)}\simeq
-0.4$, the split configuration evolves in time with slight
distortions, but it preserves the qualitative features of the
initial state for all the time of the simulation (left panel of
Fig.~\ref{fig5}). When $\gamma_{12}\simeq
\gamma_{\mathrm{cr}}^{(2)}$, the energetic instability of the
soliton pair in $BB_2$ is reflected by a \textit{dynamical}
instability: the second-species density distribution is gradually
attracted by the first species (right panel of Fig.~\ref{fig5}),
ending with an overlapped configuration, which is eventually
stabilized by radiating wave packets \cite{Yulin,garnier}.

\begin{figure}
\vskip -0.75cm \centerline{
\includegraphics[width=4.8cm]{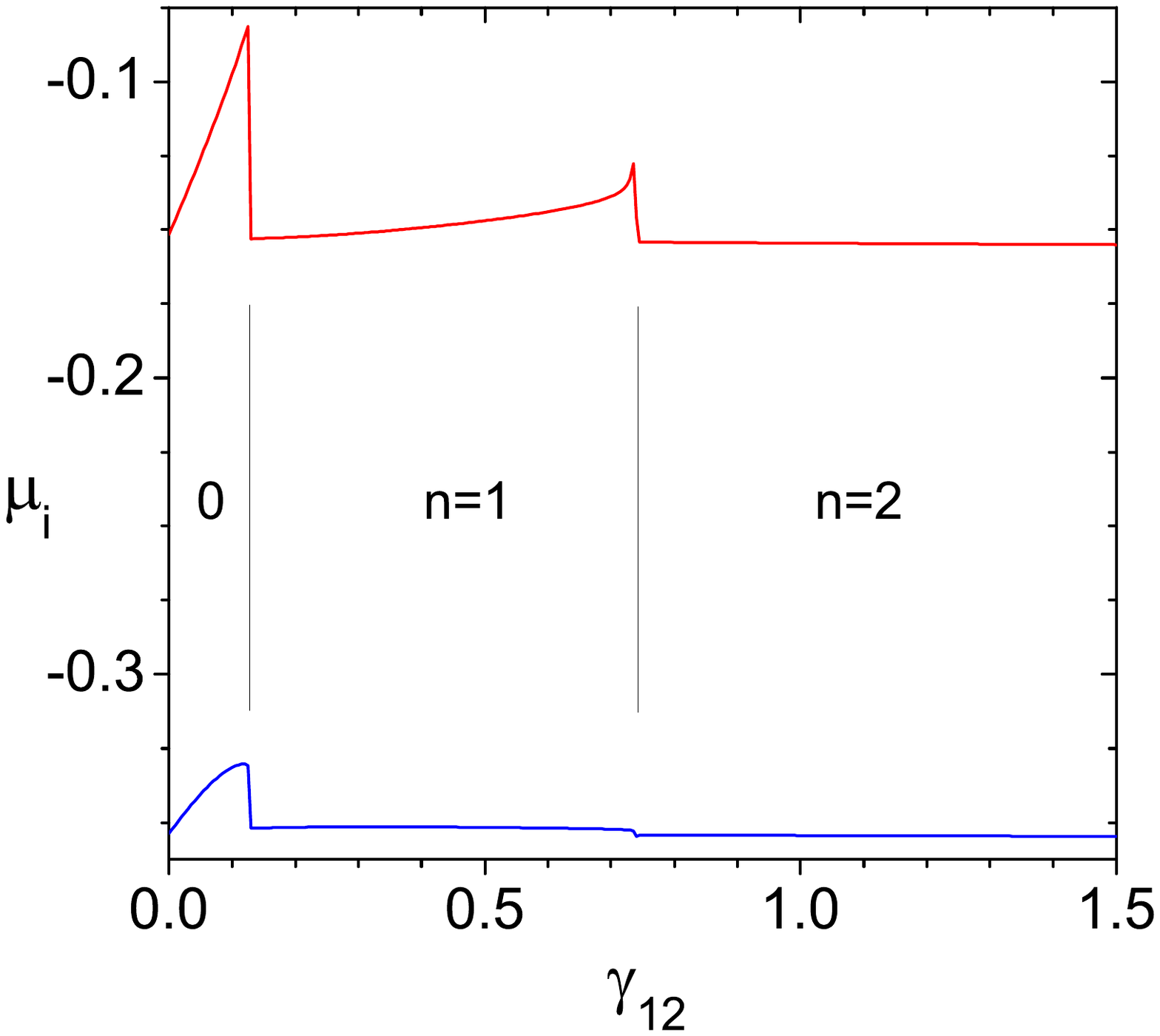}
\hskip -0.5cm
\includegraphics[width=4.8cm]{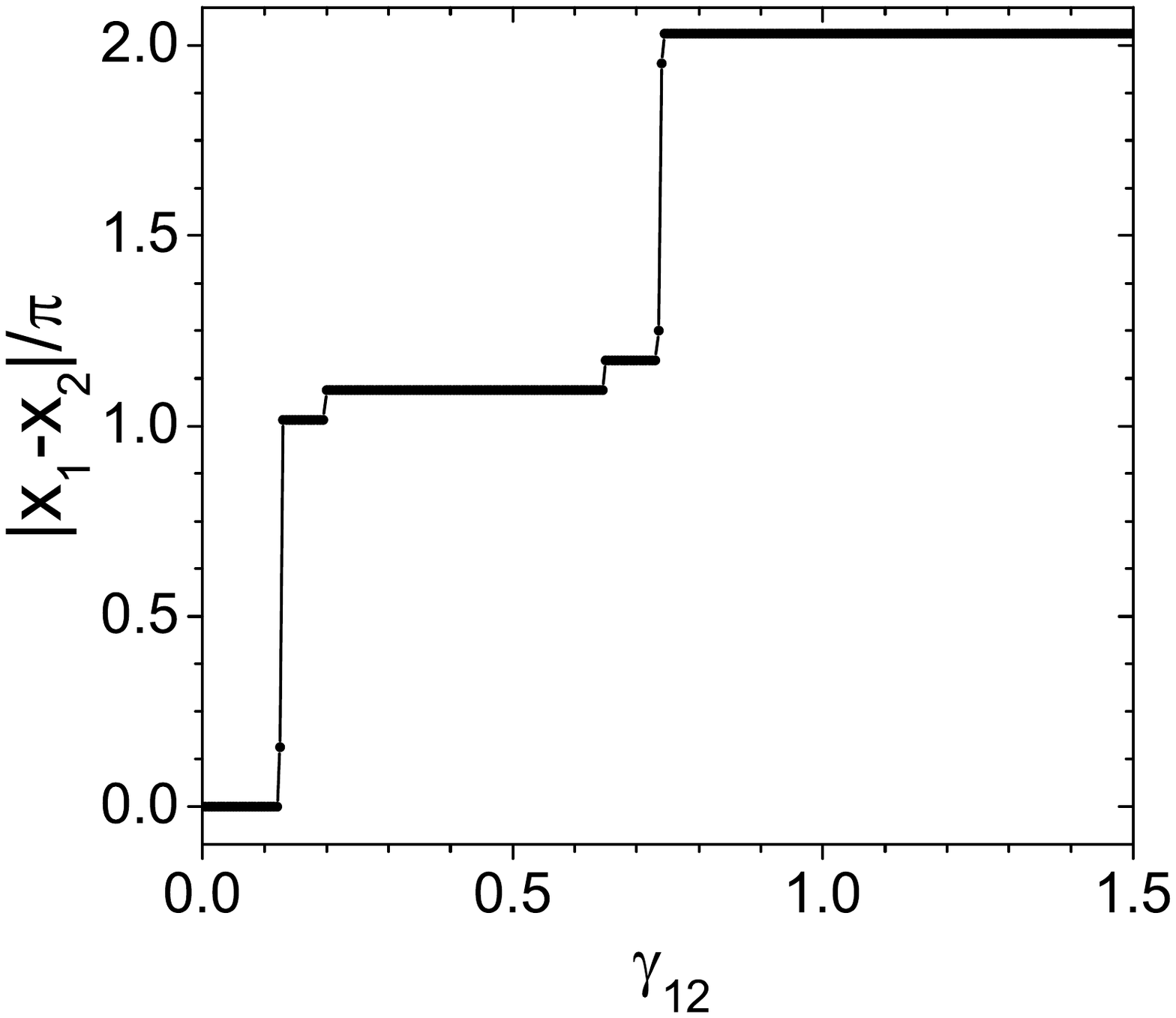}
} \vskip -2.1cm \caption{Dependence of chemical potentials (left panel) and
distance between peaks (right panel) on repulsive
interspecies interactions of a BB soliton with
$N_1=N_2=1$. Other parameters are fixed as $\gamma_1=\gamma_2=-1,
V_{01}=-0.5, V_{02} =-0.25$. Top (blue) and bottom (red) curves in the left panel
refer to  chemical potentials of first and second component, respectively, while
vertical lines separate $\gamma_{12}$ regions for existence of BB solitons in
$BB_0$ (overlapped), $BB_1$ and $BB_2$ (split). }
\label{fig3ms}
\end{figure}

\section{Splitting dependence on interspecies interaction}

In the previous section we observed that split BB
solitons can become unstable at some negative critical values of
the interspecies scattering length. We shall now investigate these
critical values in more detail by direct numerical integration of
the GPEs, both for attractive and repulsive interatomic
interactions.

\begin{figure}
\vskip -0.75cm  \centerline{
\includegraphics[width=4.8cm]{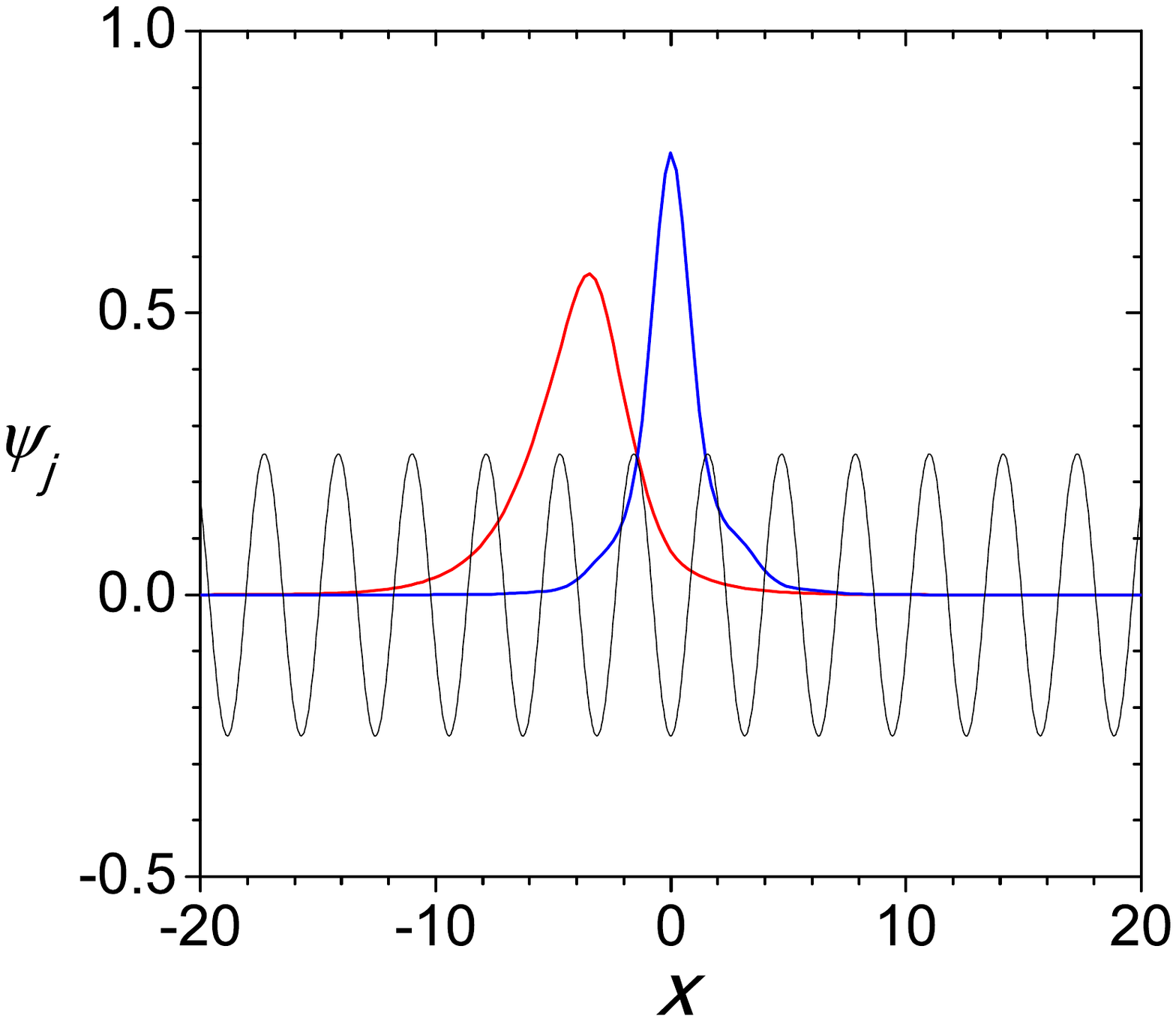}
\hskip -0.5cm
\includegraphics[width=4.8cm]{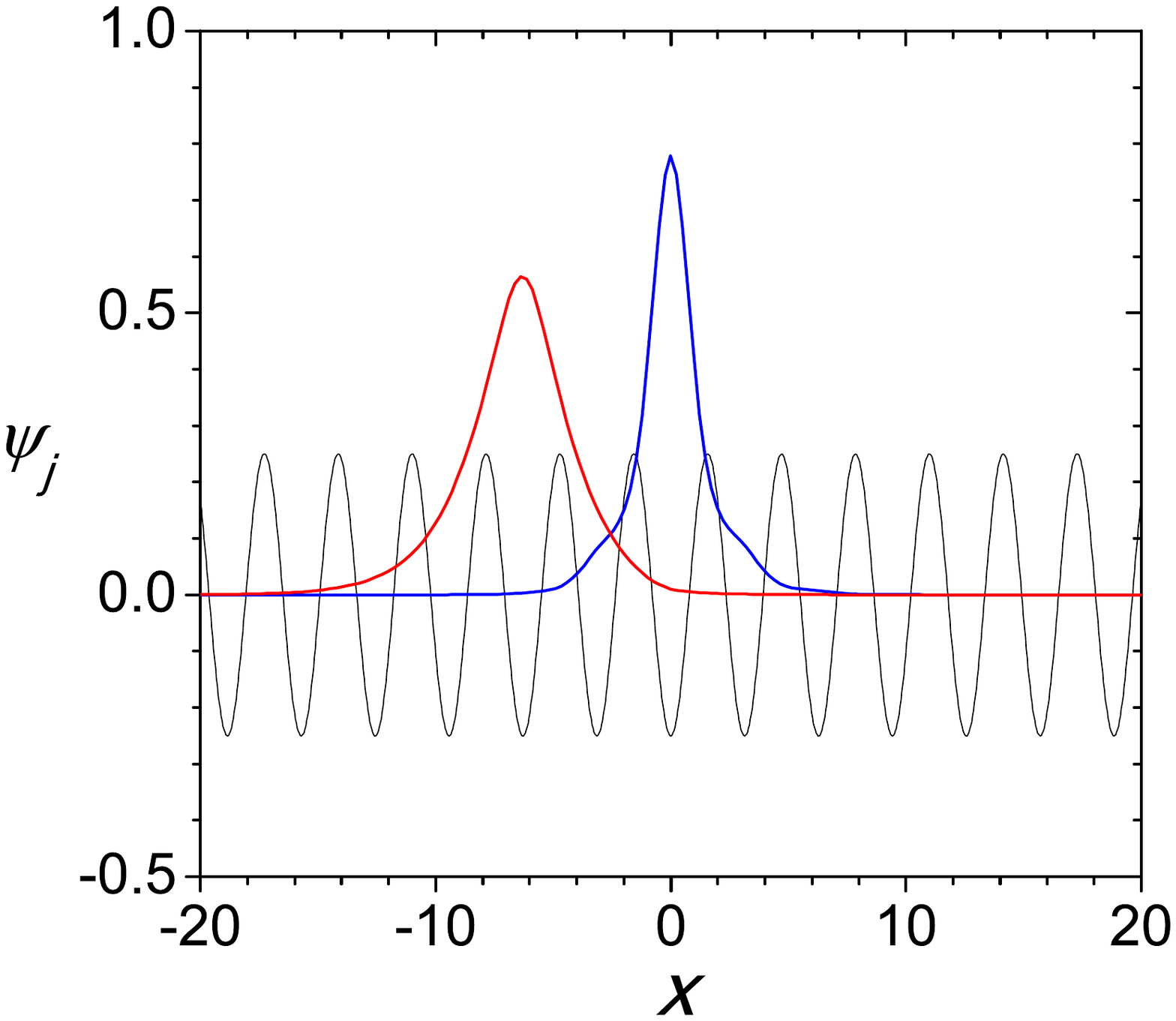}
} \vskip -2.1cm \caption{Split BB solitons inside
the regions $n=1$ and $n=2$ of the left panel of Fig.~\ref{fig3ms}, at
$\gamma_{12}=0.5$ (left panel) and $\gamma_{12}=1.0$ (right
panel). The blue (curve centered at zero) and red profiles refer to first and second component, respectively,  while the black line shows the periodicity of the  optical lattice.}\label{fig4ms}
\end{figure}

Let us first discuss the repulsive case. We can
consider as  initial state an overlapped BB soliton, centered
at $x=0$, with no interspecies coupling $(\gamma_{12}=0)$, and
adiabatically switch on a repulsive interspecies interaction
between components at $t>0$. One expects that, due to the
repulsive interspecies interaction, the initial $BB_0$ soliton
will evolve into a split one belonging to the $BB_1$ family at
some value $\gamma_{12}=\gamma_{\mathrm{rep}}^{(1)}$ and then into
the $BB_2$ family at $\gamma_{12}=\gamma_{\mathrm{rep}}^{(2)}$. This
picture coincides with the numerical results in Fig.~\ref{fig3ms}, both in terms of the chemical potentials and
the distances between peaks $\Delta x$, normalized to $\pi$. The jumps in the
distance are correlated with jumps in the chemical potentials at
the critical values, which are uniquely fixed by the parameters of
the system. In Fig.~\ref{fig4ms}, the profiles of the split
solitons with $n=1$ and $n=2$ are represented, at two different
$\gamma_{12}$ values belonging to their existence curve. Despite the smaller value of $\gamma_{12}$, the two BB components in the $n=1$ case appear to be more distorted in their overlapping region
than in the $n=2$ case. This is an evident consequence of the
exponential decay of the soliton-soliton interaction with distance [see Eq.~(\ref{eq11})]. Notice that, as in the attractive
case, the normalized distance between soliton centers is not an
integer number. This is a clear consequence of the existence of a
repulsive force between components.

\begin{figure}
\vskip -0.75cm \centerline{
\includegraphics[width=4.8cm]{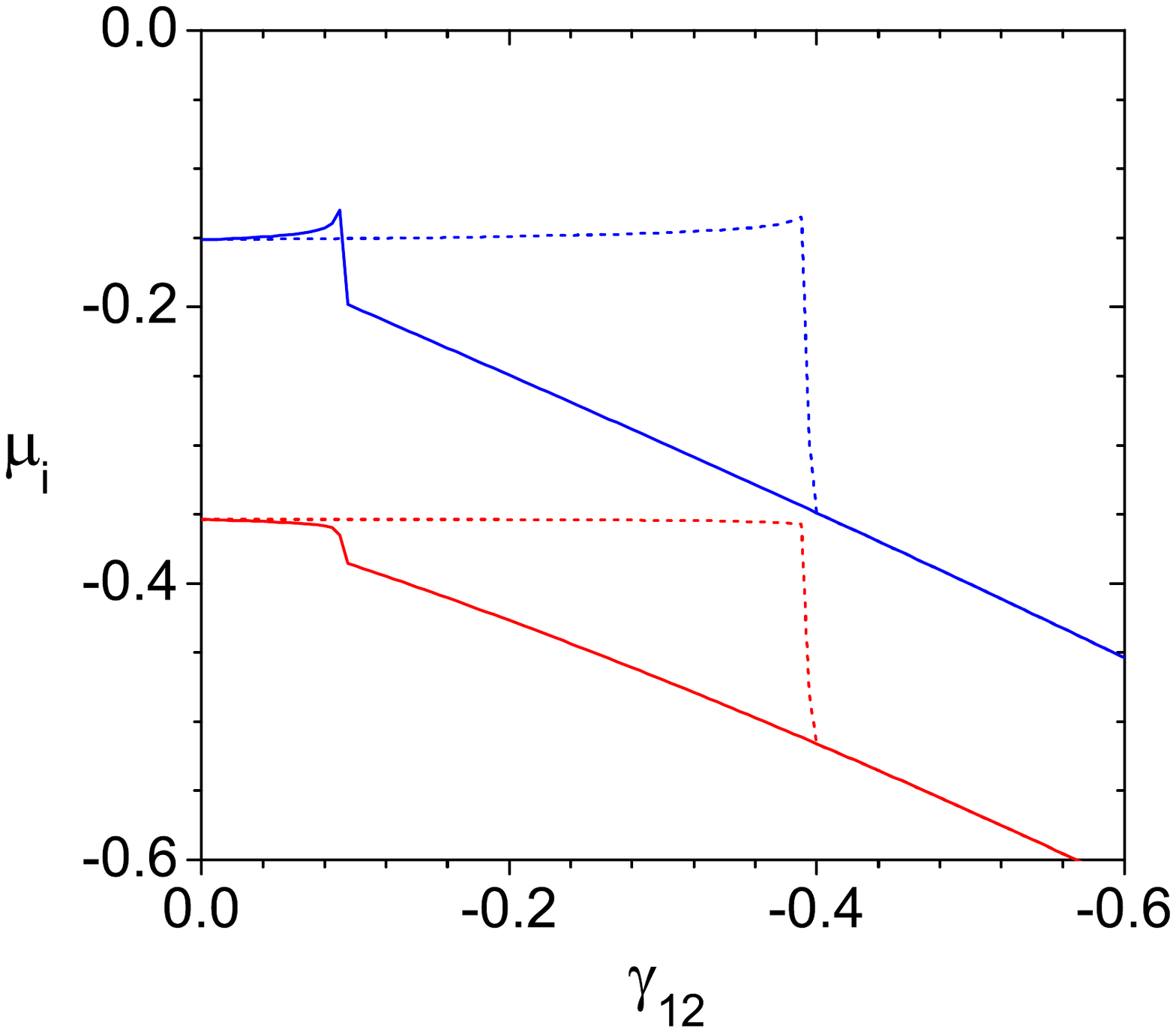}
\hskip -0.5cm
\includegraphics[width=4.8cm]{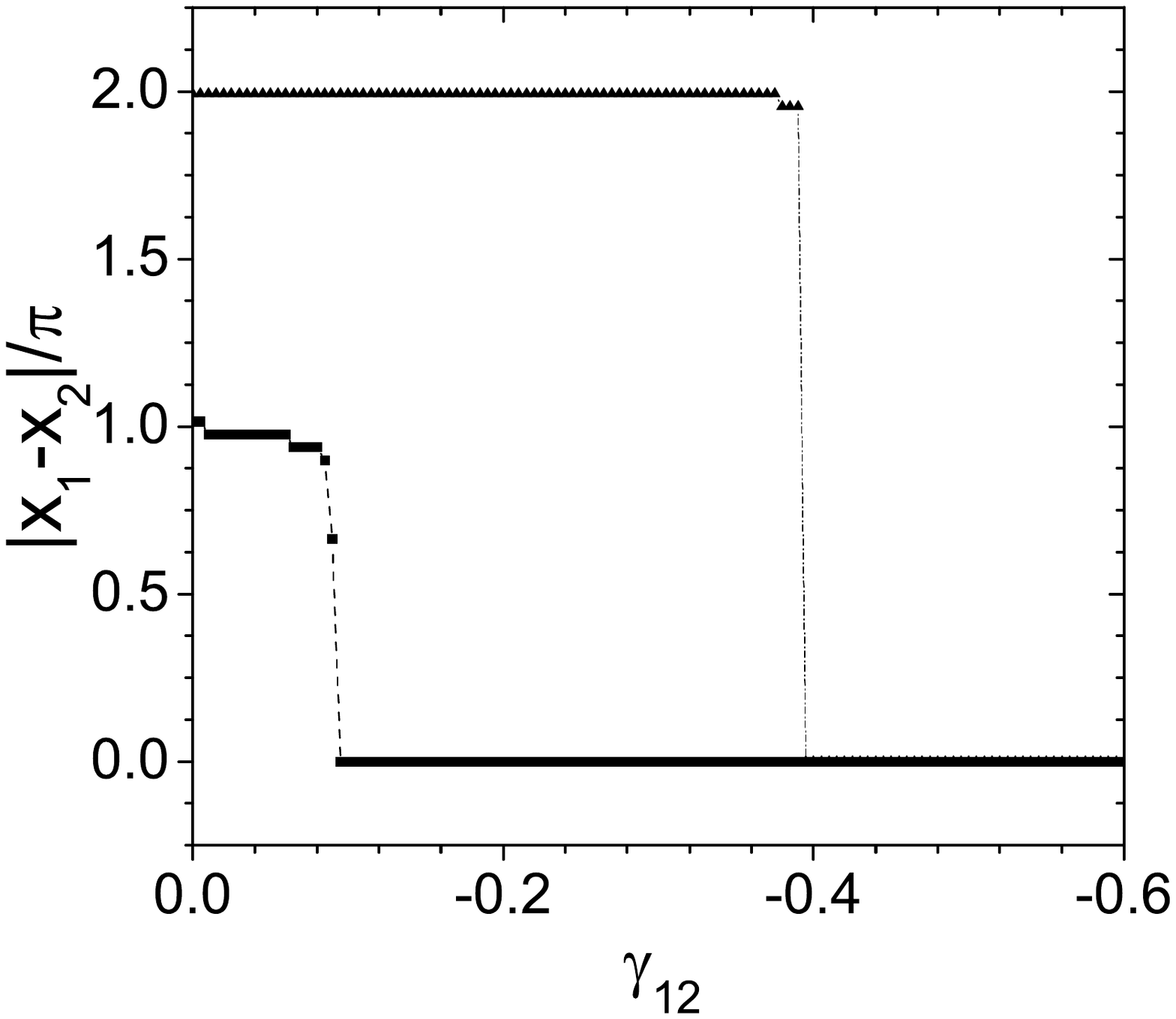}
}
\vskip -2.1cm
\caption{Dependence on attractive interspecies interaction of chemical potentials (left panel) and  distance between density centers, $x_1, x_2$, (right panel) of a BB soliton with  $N_1=N_2=1$.
Bottom (blue) and top (red) curves in the left panel  refer to the first and the second component, respectively, while dotted and continuous lines (left panel) and square and triangle symbols (right panel) refer to BB solitons with centers initially separated by $\pi$ and by $2 \pi$, respectively.
Other parameters are fixed as $\gamma_1=\gamma_2=-1,  V_{01}=-0.5, V_{02} =-0.25$.
}
\label{fig5ms}
\end{figure}

When attractive interactions $\gamma_{12}<0$ are
adiabatically turned on at $t>0$, we expect that an initial split
soliton in $BB_{n_0}$, with $n_0>0$, will undergo \textit{only one
jump} towards $n=0$. Indeed, from the analysis in the previous
section, we can deduce that the negative critical values are
ordered as $\gamma_{\mathrm{cr}}^{(n)} >
\gamma_{\mathrm{cr}}^{(n+1)}$. Thus, if $\gamma_{12} >
\gamma_{\mathrm{cr}}^{(n_0)}$, interactions give enough energy to
overcome all the intermediate barriers from the $n_0$-th down to
$n=0$. This intuitive result, based on energetic considerations,
match very well the results of the numerical simulation, as
one can see from Fig.~\ref{fig5ms}, where the cases $n_0=2$ and
$n_0=1$ are represented.

Since the critical values of $\gamma_{12}$  at which
the transitions occur are uniquely fixed by the parameters of the
mixture, including the number of atoms and intraspecies interactions,
an experimental implementation of the above numerical simulations
could be used for indirect measurements of the interspecies
scattering length of BEC mixtures. The interspecies scattering
length can also be measured from the oscillatory motion of coupled
solitons as predicted in \cite{sekh}.

\section{Conclusions}

We have considered matter-wave bright-bright solitons in coupled
Bose-Einstein condensates, by assuming that the first component is
loaded in a linear optical lattice and the second component in a
nonlinear optical lattice. In particular, the existence and stability of split and overlapped BB solitons has been  investigated  by VA, by direct numerical integrations of the coupled GPEs, and by direct numerical integrations of the system. The  dependence of the existence ranges of BB solitons on the  interspecies  interaction parameter has been also investigated. In particular, for repulsive interspecies interactions we showed the existence of a series of critical   values of $\gamma_{12}$ at which transitions  from  the  $n$- to the $n + 1$- split BB soliton occur. For attractive interspecies interaction we showed that only direct transitions from  a split BB solitons to the overlapped BB soliton are possible. Since critical values at which transitions occur depend on physical parameters of the mixture, these phenomena  suggest that split BB solitons  could be used for  indirect measurements of these parameters in  experiments.

\section*{Acknowledgements}

G. A. Sekh is thankful to INFN, Italy for providing a Post
Doctoral Fellowship and University of Kashmir, India for giving a
without-pay-leave to enjoy the fellowship. G. A. Sekh is grateful
to Benoy Talukdar for useful discussions. M. S. acknowledges
partial support from the Ministero dell'Istruzione,
dell'Universit\`a e della Ricerca (MIUR) through a PRIN (Programmi
di Ricerca Scientifica di Rilevante Interesse Nazionale) 2010-2011
initiative. P. F. is partially supported by the Italian National Group of Mathematical Physics (GNFM-INdAM).
P. F., F. V. P. and S. P. are partially supported by the PRIN Grant No.\ 2010LLKJBX on ``Collective quantum phenomena: from strongly correlated systems to quantum simulators".

\end{document}